\documentclass{article}
\usepackage{dcolumn}
\usepackage{bm}
\usepackage{latexsym}
\usepackage{extarrows}
\usepackage{amsmath}
\usepackage{graphicx}
\usepackage{amsmath,amssymb}
\usepackage{ulem}
\usepackage{authblk}
\usepackage[colorlinks,linkcolor=blue, urlcolor=blue ,anchorcolor=blue,citecolor=green]{hyperref}
\newcommand{\be}{\begin{equation}}
\newcommand{\ee}{\end{equation}}
\newcommand{\bea}{\begin{eqnarray}}
\newcommand{\eea}{\end{eqnarray}}

\title{Anisotropic Inflation with General Potentials}

\author{Jiaming Shi$^1$}

\author{Xiaotian Huang$^1$}

\author{Taotao Qiu$^{1,2}$ \footnote{qiutt@mail.ccnu.edu.cn}}

\affil{\scriptsize{$1$ Department of Physics Science and Technology, Central China Normal University, Wuhan, 430079, China}}
\affil{\scriptsize{$2$ Institute of Astrophysics, Central China Normal University, Wuhan 430079, China}}
\begin{document}

\maketitle

\begin{abstract}
Anomalies in recent observational data indicate that there might be some ``anisotropic hair" generated in an inflation period. To obtain general information about the effects of this anisotropic hair to inflation models, we studied anisotropic inflation models that involve one vector and one scalar using several types of potentials. We determined the general relationship between the degree of anisotropy and the fraction of the vector and scalar fields, and concluded that the anisotropies behave independently of the potentials. We also generalized our study to the case of multi-directional anisotropies.
\text{PACS numbers: 98.80.Cq, 98.80.Es, 03.50.De}
\end{abstract}

\section{Introduction}
Inflation \cite{Guth:1980zm,Linde:1981mu,Starobinsky:1980te} is considered as one of the most successful theories in modern cosmology. By almost preserving de-Sitter symmetry in the universe, it can not only solve the horizon, flatness, and unwanted relics problems, but can also  generate a scale-invariant primordial power spectrum, which fits the observational data provided by the Cosmic Microwave Background (CMB). After being introduced in the 1980s, thousands of inflation models have been proposed, with various motivations and different observational features \cite{Martin:2013tda}. Although recent observations have been imposing increasingly  tighter constraints on those models, it is still difficult to determine which model is the best.

Recently, the PLANCK satellite, which is sponsored by the European Space Agency, released its second-year data on CMB. Although a homogeneous and isotropic universe is still strongly favored, the newly-released data also implies that some "anomalies" might exist, such as the lack of large-scale correlations, asymmetry between northern and southern hemispheres, various parity symmetries, and cold spots. Some of these anomalies indicate that there might be a degree of directionality, or ``anisotropic hair", in the early universe. It would therefore be interesting to model the universe using anisotropic matter, in order to understand its physical origin.

It is assumed that the anisotropic hair is caused by some vector fields during inflation, which is then called "anisotropic inflation". Such a viable model has been proposed in \cite{Watanabe:2009ct} (see also \cite{Hervik:2011xm}), in which the inflaton field $\phi$ has the simplest mass squared term, whereas the coupling terms of the vector field to the inflaton has the form $f^2(\phi)F_{\mu\nu}F^{\mu\nu}$ with an arbitrary function $f(\phi)$. In \cite{Watanabe:2009ct}, it was determined that an attractor solution of $R\equiv\rho_A/\rho_\phi\sim 10^{-2}$ can be obtained for various parameter choices, where $\rho_A$ and $\rho_\phi$ denote the energy density of the vector and the inflaton, respectively. Moreover, the degree of anisotropy, $\Sigma\equiv\dot\sigma/H$, is proportional to $R$. Interestingly, these results imply that the anisotropy caused by the vector field does have an effect on the inflation but without violating global isotropy. Subsequently, anisotropic inflation attracted and more attention in the literature. For instance, Refs. \cite{Kanno:2010nr,Emami:2010rm,Murata:2011wv,Ohashi:2013pca} studied anisotropic inflation with various types of scalar and vector fields, and Refs. \cite{Dulaney:2010sq,Gumrukcuoglu:2010yc,Watanabe:2010fh,Bartolo:2012sd,Emami:2013bk,Abolhasani:2013zya,Ohashi:2013qba} discussed relevant perturbation theories. More comprehensive investigations are presented in the review of Ref. \cite{Maleknejad:2012fw} and, very recently, Ref. \cite{Chen:2014zoa} examined how the anisotropies are affected in the presence of non-minimal gravitational coupling.

Although the analysis in \cite{Watanabe:2009ct} seems to depend on the specific form of the inflaton potential and function $f(\phi)$, it can be extended to more general potentials to investigate whether the vector field is applicable to different inflation models, which is the aim of our work. The rest of the paper is organized as follows: Sec. 2 provides the general formulae for the vector inflation model without a specific potential form. In Sec. 3, we classify the general potentials into four types and analyze the anisotropy effect for each case; analytical and numerical calculations are provided. In Sec. 4, we extend our study to the case of multi-directional anisotropy, and Sec. 5 presents the conclusions.

\section{Anisotropic Inflation: Basic Formulae}
We start with a general anisotropic inflation action, which was first proposed in \cite{Watanabe:2009ct}:
\be\label{action}
S=\int d^4x\sqrt{-g}\left[\frac{R}{2\kappa^2}-\frac{1}{2}\partial_\mu\phi\partial^\mu\phi-V(\phi)-\frac{1}{4}f^2(\phi)F_{\mu\nu}F^{\mu\nu}\right]~,
\ee
where $g$ is the determinant of metric, $R$ is the Ricci scalar, $\phi$ is the scalar field acting as the inflaton with a potential of the general form $V(\phi)$, and $F_{\mu\nu}\equiv\partial_\mu A_\nu-\partial_\nu A_\mu$ with $A_\mu$ being the vector field. Moreover, we set $\kappa^2=8\pi G$ with $G$ being the gravitational constant. From action (\ref{action}), we can see that the vector field couples with the inflaton via an arbitrary function $f(\phi)$. Owing to the presence of the vector field which is expected to provide the anisotropic hair, the exact isotropic Friedmann-Robertson-Walker (FRW) metric is not applicable. Throughout this work, we consider the simplest anisotropic extension, namely, the axisymmetric Bianchi type I metric of the form
\be\label{metric}
ds^2=-dt^2+e^{2\alpha(t)}\left[e^{-4\sigma(t)}dx^2+e^{2\sigma(t)}(dy^2+dz^2)\right]~,
\ee
where $t$ is the cosmic time, $e^{2\alpha(t)}$ is the isotropic scale factor, and $\sigma$ represents a deviation from isotropy. Here for convenience and without loss of generality, we assume that the anisotropic hair is oriented along the $x$-axis, and the vector field $A_\mu$ is also assumed to be $(0,A_x(t),0,0)$; therefore, the plane perpendicular to the vector exhibits plane symmetry.

By varying action (\ref{action}) with respect to $\phi$ and $A_\mu$, we obtain the equations of motion for the scalar and vector field as
\be\label{Aeom}
\dot{A_x}=q_0f^{-2}(\phi)e^{-\alpha-4\sigma}~,
\ee
where the over-dot denotes the derivative with respect to $t$, $q_0$ is an integration constant, and
\be\label{phieom}
\ddot{\phi}=-3\dot{\alpha}\dot{\phi}-V'(\phi)+q_0^2f^{-3}(\phi)f'(\phi)e^{-4\alpha-4\sigma}~,
\ee
where the prime denotes the derivative with respect to $\phi$. Moreover, using the Einstein equations we obtain the following Friedmann equations
\bea
\label{Fr1}
\dot{\alpha}^2&=&\dot{\sigma}^2+\frac{\kappa^2}{3}\left[\frac{1}{2}\dot{\phi}^2+V(\phi)+\frac{q_0^2}{2}f^{-2}(\phi)e^{-4\alpha-4\sigma}\right]~,\\
\label{Fr2}
\ddot{\alpha}&=&-3\dot{\alpha}^2+\kappa^2V(\phi)+\kappa^2\frac{q_0^2}{6}f^{-2}(\phi)e^{-4\alpha-4\sigma}~,\\
\label{ani}
\ddot{\sigma}&=&-3\dot{\alpha}\dot{\sigma}+\kappa^2\frac{q_0^2}{3}f^{-2}(\phi)e^{-4\alpha-4\sigma}~,
\eea
where the Hubble parameter can be expressed as $H\equiv\dot\alpha$. From (\ref{action}), the energy density and pressure of $\phi$ and $A_\mu$ are
\be
\rho_\phi=\frac{1}{2}\dot{\phi}^2+V~,~p_\phi=\frac{1}{2}\dot{\phi}^2-V~,~\rho_A=
\frac{q_0^2}{2}f^{-2}(\phi)e^{-4\alpha-4\sigma}~,~p_A=
\frac{q_0^2}{6}f^{-2}(\phi)e^{-4\alpha-4\sigma}~
\ee
respectively. This indicates that the equation of state, $w\equiv p/\rho$, of $\phi$ and $A_\mu$ are $w_\phi=(\dot{\phi}^2-2V)/(\dot{\phi}^2+2V)$ and $w_A=1/3$ respectively, consistent with the normal canonical inflaton field and radiation.

In \cite{Watanabe:2009ct}, it is assumed that the total energy density is dominated by the potential form of the inflaton; namely, the conditions ${\dot\sigma^2, \rho_A, \dot\phi^2/2}\ll V(\phi)$ have been imposed. Eqs. (\ref{phieom}) and (\ref{Fr1}) become
\be\label{phieomFr1SR}
\dot{\alpha}^2\simeq\frac{\kappa^2}{3}V(\phi)~,~~~3\dot{\alpha}\dot{\phi}\simeq-V'(\phi)~,
\ee
which provide the solution
\be\label{alphadotphi}
\alpha(\phi)\simeq-\kappa^2\int_{\phi_i}^\phi\frac{V}{V'}d\phi~,~~~\dot\phi\simeq-\frac{V'}{\kappa\sqrt{3V}}~.
\ee
Assuming
\be\label{f}
f(\phi)=e^{2c\kappa^2\int\frac{V}{V'}d\phi}
\ee
where $c$ is a parameter, and considering the equation of motion for $\phi$ and the vector field, we have:
\be\label{phieomSR2}
3\dot{\alpha}\dot{\phi}\simeq-V'(\phi)+ 2c\kappa^2q_0^2\frac{V}{V'}e^{-4c\kappa^2\int(V/V')d\phi-4\alpha-4\sigma}~,
\ee
or (by using (\ref{phieomFr1SR}))
\be\label{phieomSR3}
\frac{d\phi}{d\alpha}=-\frac{V'}{\kappa^2V}+\frac{2cq_0^2}{V'}e^{-4c\kappa^2\int(V/V')d\phi-4\alpha-4\sigma}~.
\ee

We call the time when the last term on the right-hand side of Eq. (\ref{phieomSR3}) begins to dominate over the first term "phase transition"; after that time, Eq. (\ref{phieomSR3}) can be solved to obtain
\be\label{solA}
f^{-2}e^{-4\alpha-4\sigma}\simeq\frac{(c-1)V'^2}{2c^2\kappa^2q_0^2V}
\ee
for $c>1$.
By defining the energy density ratio between the scalar and vector fields ${\cal R}$ and the degree of anisotropy ${\cal E}$ as:
\be\label{ratiodegree}
{\cal R}\equiv\frac{\rho_A}{\rho_\phi}=\frac{p_A^2f^{-2}(\phi)e^{-4\alpha-4\sigma}}{\dot{\phi}^2+2V}~,~~~{\cal E}\equiv\frac{\dot{\sigma}}{\dot{\alpha}}~,
\ee
we can use Eqs. (\ref{ani}), (\ref{phieomFr1SR}), and (\ref{solA}),and the slow-roll condition $\dot{\phi}^2/2\ll V$ to derive:
\be\label{ER}
{\cal R}=\frac{c-1}{4c^2\kappa^2}\left(\frac{V'}{V}\right)^2~,~~~{\cal E}=\frac{2}{3}{\cal R}~.
\ee
In Eq. (\ref{ani}), $\ddot\sigma$ is also neglected. Moreover, from Eq.s (\ref{Fr1}), (\ref{Fr2}), (\ref{phieomFr1SR}), (\ref{phieomSR2}) and (\ref{solA}) we can obtain the slow-roll parameters:
\be\label{epsilon}
\epsilon_\alpha\equiv-\frac{\ddot\alpha}{\dot\alpha^2}=\frac{1}{2c\kappa^2}\left(\frac{V'}{V}\right)^2~,~~~\epsilon_V\equiv\frac{1}{2\kappa^2}\left(\frac{V'}{V}\right)^2~,
\ee
so the relationship between ${\cal R}$ and $\epsilon_\alpha$, $\epsilon_V$ can be obtain as:
\be\label{Repsilon}
{\cal R}=\frac{c-1}{2c}\epsilon_\alpha=\frac{c-1}{2c^2}\epsilon_V~.
\ee
A similar analysis is also provided in e.g., \cite{Ohashi:2013qba} and \cite{Chen:2014zoa}.

\begin{figure}[htbp]
\centering
\includegraphics[width=2.50in,height=1.80in]{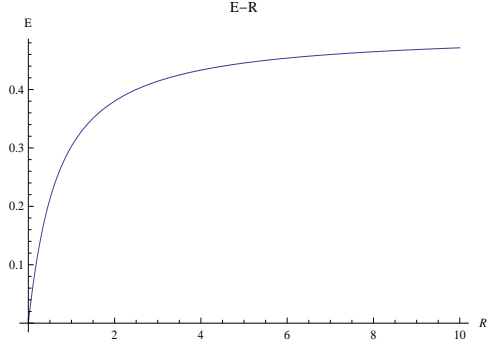}\\
\caption{The relation between ${\cal E}$ and ${\cal R}$.}
\label{ERplot}
\end{figure}

In this study, we extend to the more general case, in which $\dot\sigma^2$ and $\rho_A$ are not considered negligible compared to $V$, to investigate whether their presence alters the relations (\ref{ER}) and (\ref{Repsilon}). In this case, Eq. (\ref{Fr1}) is applied instead of Eq. (\ref{phieomFr1SR}). First, using Eqs. (\ref{Fr1}) and (\ref{ani}) as well as the definition of ${\cal R}$ and ${\cal E}$, we obtain
\be\label{Fr1ani}
\dot{\alpha}^2=\frac{2\kappa^2}{9}{\cal R}{\cal E}V+\frac{\kappa^2}{3}({\cal R}+1)V~,~~~\dot{\alpha}^2\simeq\frac{2\kappa^2}{9}\frac{{\cal R}V}{{\cal E}}~,
\ee
which can be combined together to obtain a more general relation of ${\cal E}$ and ${\cal R}$:
\be\label{ER2}
{\cal E}=\frac{-3-3{\cal R}+\sqrt{9+18{\cal R}+25{\cal R}^2}}{4{\cal R}}~.
\ee
We can see that in the limit of ${\cal R}\ll 1$, Eq. (\ref{ER2}) can be expanded as ${\cal E}\simeq(2/3){\cal R}-(2/3){\cal R}^2+(10/27){\cal R}^3+o({\cal R}^4)$, and the leading order term of this equation reproduces (\ref{ER}). However, as ${\cal R}$ increases, the linear relation between ${\cal E}$ and ${\cal R}$ is no longer maintained. We have numerically plotted the relation of ${\cal E}$ between ${\cal R}$ in Fig. \ref{ERplot}; and the plot demonstrates that, when ${\cal R}$ becomes large, ${\cal E}$ approaches a constant value instead of increasing proportionally to ${\cal R}$. In our plot, we show that the maximum value that ${\cal E}$ can obtain is approximately ${\cal E}_{max}\simeq0.5$. This implies that the growth of anisotropy never exceeds that of the background, and inflation is not destroyed. However, when the vector field plays an important role during inflation, care should be taken not to neglect the part containing the anisotropy terms in the total energy density.

Moreover, Eqs. (\ref{epsilon}) and (\ref{Repsilon}) are also modified if we consider the ratio of the vector field and anisotropy to the total energy density. By defining the energy density fraction of the inflaton as $\Omega_\phi\equiv\kappa^2\rho_\phi/(3\dot\alpha)^2$, the first equation of Eq. (\ref{Fr1ani}) can be written as:
\be\label{Fr1ani2}
\dot{\alpha}^2=\frac{\kappa^2}{3}\frac{V}{\Omega_\phi}~,~~~\Omega_\phi=\frac{1-{\cal E}^2}{1+{\cal R}}~.
\ee
thus, Eq.(\ref{phieomSR3}) is modified as
\be\label{phieomFr1SR4}
\frac{d\phi}{d\alpha}=-\Omega_\phi\frac{V'}{\kappa^2V}+\Omega_\phi\frac{2cq_0^2}{V'}e^{-4c\kappa^2\int V/V'd\phi-4\alpha-4\sigma}~,
\ee
with the solution:
\be\label{solA2}
f^{-2}e^{-4\alpha-4\sigma}\simeq\frac{(c\Omega_\phi-1)V'^2}{2c^2\kappa^2q_0^2\Omega_\phi V}
\ee
for $c>1$.

From the definition of ${\cal R}$ in (\ref{ratiodegree}), we can obtain the modified expression of ${\cal R}$, which is:
\be\label{R2}
{\cal R}=\frac{c\Omega_\phi-1}{4c^2\kappa^2\Omega_\phi}\left(\frac{V'}{V}\right)^2~,
\ee
so we can derive the relationship between ${\cal R}$ and $\epsilon_V$ as:
\be\label{R2epsilonV}
{\cal R}=\frac{c\Omega_\phi-1}{2c^2\Omega_\phi}\epsilon_V~,
\ee
and  using Eqs. (\ref{ER2}) and (\ref{Fr1ani2}) we have:
\be\label{R2epsilonV2}
{\cal R}=\frac{18c^2\epsilon_V-36c^3\epsilon_V-9c\epsilon_V^2+3\sqrt{36c^4\epsilon_V^2+36c^3\epsilon_V^3-16c\epsilon_V^4+25c^2\epsilon_V^4})}
{4(-18c^4-9c^2\epsilon_V+2\epsilon_V^2)}~.
\ee

The above analysis shows that in the limit of $\epsilon_V\ll 1$, we have ${\cal R}\simeq\frac{c-1}{2c^2}\epsilon_V-\frac{c-1}{4c^4}\epsilon_V^2+o(\epsilon_V^3)$ and the leading order of this equation is consistent with Eq. (\ref{Repsilon}). However, at later times during the inflation, when $\epsilon_V\rightarrow 1$, it will involve large corrections from the higher orders of $\epsilon_V$.

The expression of $\epsilon_\alpha$ is different from (\ref{epsilon}) because of the presence of the anisotropy and vector parts. From Eqs. (\ref{Fr1}), (\ref{Fr2}), (\ref{Fr1ani}) and (\ref{phieomFr1SR4}), we obtain:
\bea
\epsilon_\alpha&=&3{\cal E}^2+\frac{1}{2c^2\kappa^2}\left(\frac{V'}{V}\right)^2+\frac{3(c\Omega_\phi-1){\cal E}}{4c^2\kappa^2{\cal R}\Omega_\phi}\left(\frac{V'}{V}\right)^2~,\nonumber\\
&=&3{\cal E}^2+\frac{2\Omega_\phi{\cal R}}{(c\Omega_\phi-1)}+3{\cal E}~,
\eea
which can be reduced to provide the relationship of ${\cal R}$ and $\epsilon_\alpha$ as:
\be\label{R2epsilonalpha}
{\cal R}\simeq\frac{c-1}{2c}\epsilon_\alpha
+\left(\frac{1}{12}-\frac{1}{4c}+\frac{1}{6c^3}\right)\epsilon_\alpha^2~,
\ee
where Eqs. (\ref{ER2}) and (\ref{Fr1ani2}) have been applied.
\begin{figure}[htbp]
\centering
\includegraphics[width=1.70in,height=1.30in]{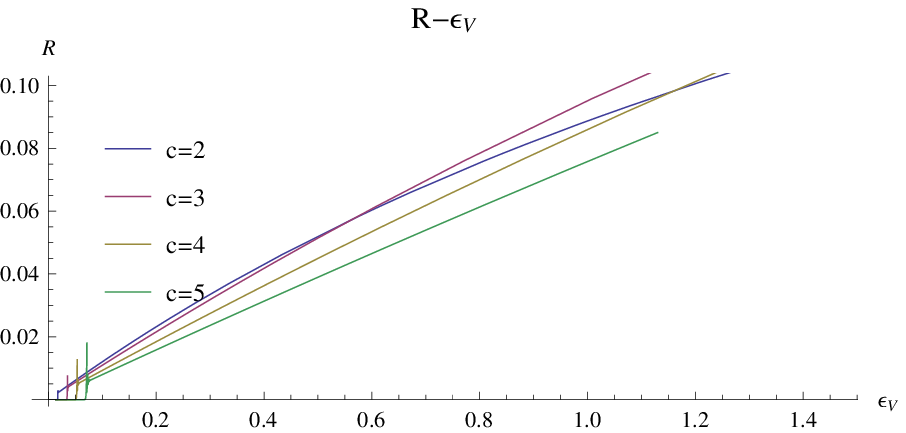}
\includegraphics[width=1.70in,height=1.30in]{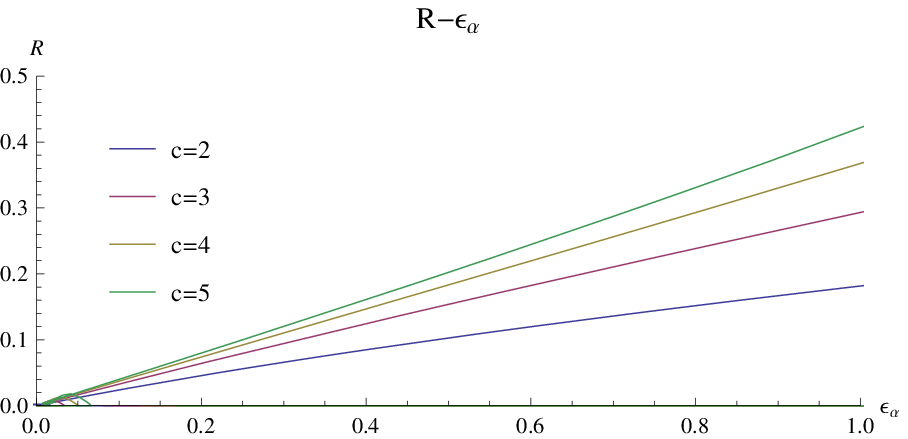}\\
\caption{The relation of $\epsilon_V$ and ${\cal R}$, $\epsilon_\alpha$ and ${\cal R}$}
\label{Repsilonplot}
\end{figure}
We have numerically plotted ${\cal R}$ in terms of $\epsilon_V$ and $\epsilon_\alpha$ according to Eqs. (\ref{R2epsilonV}) and (\ref{R2epsilonalpha}),respectively,in Fig.\ref{Repsilonplot}. From the plot, we can see that again ${\cal R}$ monotonically (but not linearly) increases with the slow-roll parameters, as it receives corrections from the higher-order terms of $\epsilon_V(\epsilon_\alpha)$. Moreover, the maximum value of ${\cal R}$ , which is achieved when the slow-roll parameters reach unity (indicating the end of the inflation), is again near ${\cal O}(0.1)$, namely, much smaller than unity. Thus, it is reasonable to assume the ${\cal R}\ll 1$ limit.

Furthermore, we obtained other useful relations. Using Eq. (\ref{phieomSR2}), (\ref{Fr1ani2}), and (\ref{solA2}), we have
\be\label{alphadotphi2}
\alpha(\phi)\simeq\alpha(\phi_T)-c\kappa^2\int_{\phi_T}^\phi\frac{V}{V'}d\phi~,~~~\dot{\phi}\simeq-\frac{V'}{c\kappa\sqrt{3\Omega_\phi V}}~.
\ee

In (\ref{alphadotphi2}), $\phi_T$ is the value of $\phi$ at the time when the phase transition occurs, which can be obtained by solving the equation:
\be\label{phiT}
-4c\kappa^2\int \frac{V}{V'}d\phi\Bigg|_{\phi=\phi_T}+4\int_{\phi_i}^{\phi_T}\frac{V}{V'}d\phi=\mathrm{ln}\left(\frac{V'^2(\phi_T)}{2c\kappa^2q_0^2V(\phi_T)}\right)
\ee


\section{Anisotropic Inflation: Explicit Examples}
In this section, we analyze anisotropic inflation more specifically. Because every single canonical scalar field model with a general monotonic potential can have either positive or negative $V'$, which actually correspond to a {\it large-field model} ($\phi$ evolves from a large value to a small one, $\dot\phi>0$) and a {\it small-field model} ($\phi$ evolves from a small value to a large one, $\dot\phi<0$), respectively. Moreover, we can also divide each case into two cases according to the sign of $V''$; namely, $V''>0$ ({\it -concave model}) or $V''<0$ ({\it -convex model}).Because $V''$ is not included in the background equations, it may not have a significant effect in the background evolution; however, it does affect perturbations, such as the tilt of the index of the scalar power spectrum or the amount of the tensor/scalar ratio. In the following, we examine the evolution of the inflation field and the effects of the vector field and anisotropy on inflation for all these cases using explicit examples.
\subsection{Analytical calculations}
\subsubsection{Large-field concave model}
\begin{figure}[htbp]
\centering
\includegraphics[width=1.70in,height=1.10in]{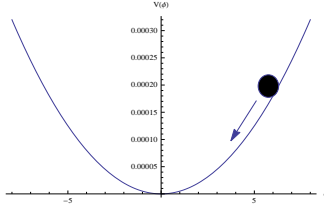}\\
\caption{The plot of $V_1(\phi)=\frac{1}{2}m^2\phi^2$.}
\label{v1plot}
\end{figure}

In this case, we have $V'>0$ and $V''>0$ for the potential $V(\phi)$. There are many well-known models that address this case; for example, the potential can be of the form:
\be\label{v1}
V_1(\phi)=\frac{1}{2}m^2\phi^2~,
\ee
which is known as the chaotic inflation \cite{Linde:1981mu}. Here, $m$ is the mass of the inflaton and, according to observational data, we have $m\sim 10^{-6}M_p$. The shape of the potential is shown in Fig. \ref{v1plot}.

From (\ref{v1}), we have
\be\label{Repsilon1}
\epsilon_{V1}=\frac{2}{\kappa^2\phi^2}~,~~~{\cal R}_1=\frac{c\Omega_\phi-1}{c^2\kappa^2\Omega_\phi\phi^2}~.
\ee
Moreover, from (\ref{alphadotphi}), (\ref{alphadotphi2}), and (\ref{phiT}), we can also obtain the efolding number $\alpha$ and the velocity of $\phi$ as functions of $\phi$,
\bea\label{alpha1}
\alpha_1(\phi)&=&
\begin{cases}
\frac{\kappa^2}{4}(\phi_i^2-\phi^2)~&(\phi\geq\phi_T)\\
\alpha_1(\phi_T)+\frac{c\kappa^2}{4}(\phi_T^2-\phi^2)~&(\phi\leq\phi_T)
\end{cases}~,\\
\label{dotphi1}
\dot\phi(\phi)&=&-\sqrt{\frac{2}{3}}\frac{m}{\kappa}~~(\phi\geq\phi_T)~,~-\sqrt{\frac{2}{3\Omega_\phi}}\frac{m}{c\kappa}~~(\phi\leq\phi_T)~,
\eea
and $\phi_T$ can be solved as:
\be\label{phiT1}
\phi_{T1}=\sqrt{-\frac{\phi_i^2}{c-1}-\frac{1}{(c-1)\kappa^2}\ln(\frac{m^2}{cq_0^2\kappa^2})}~.
\ee



\subsubsection{Large-field convex model}
\begin{figure}
\centering
\includegraphics[width=1.70in,height=1.10in]{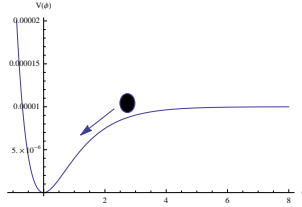}\\
\caption{The plot of $V_2(\phi)=\lambda\left(1-e^{-\mu\phi}\right)^2$.}
\label{v2plot}
\end{figure}

In this case, we have $V'>0$ and $V''<0$ for the potential $V(\phi)$. An example of this case is
\be\label{v2}
V_2(\phi)=\lambda\left(1-e^{-\mu\phi}\right)^2~.
\ee
As is well known, this potential can be obtained as the effective potential of the Starobinsky model \cite{Starobinsky:1980te,DeFelice:2010aj}, $S=(2\kappa^2)^{-1}\int d^4x\sqrt{-g}[R+\xi R^2]$, in its Einstein frame when $\lambda=(8\kappa^2\xi)^{-1}$ and $\mu=\sqrt{2/3}\kappa$, which can yield a small tensor/scalar ratio \cite{Ade:2015lrj}. The shape of the potential is displayed in Fig. \ref{v2plot}.

From (\ref{v2}), we have
\be\label{Repsilon2}
\epsilon_{V2}=\frac{2\mu^2e^{-2\mu\phi}}{\kappa^2(1-e^{-\mu\phi})^2}~,~~~{\cal R}_2=\frac{\mu^2(c\Omega_\phi-1)e^{-2\mu\phi}}{c^2\kappa^2\Omega_\phi(1-e^{-\mu\phi})^2}~,
\ee
and the efolding number $\alpha$ and the velocity of $\phi$ can be obtain as functions of $\phi$ :
\bea\label{alpha2}
\alpha_2(\phi)&=&
\begin{cases}
\frac{\kappa^2}{2\mu^2}[e^{\mu\phi_i}-e^{\mu\phi}+\mu(\phi-\phi_i)]~&(\phi\geq\phi_T)\\
\alpha_2(\phi_T)+\frac{c\kappa^2}{2\mu^2}[e^{\mu\phi_T}-e^{\mu\phi}+\mu(\phi-\phi_T)]~&(\phi\leq\phi_T)
\end{cases}~,\\
\label{dotphi2}
\dot\phi(\phi)&=&-\frac{2\mu}{\kappa}\sqrt{\frac{\lambda}{3}}e^{-\mu\phi}~~(\phi\geq\phi_T)~,~-\frac{2\mu}{c\kappa}\sqrt{\frac{\lambda}{3\Omega_\phi}}e^{-\mu\phi}~~(\phi\leq\phi_T)~,
\eea
where $\phi_T$ is the solution of the equation:
\be\label{phiT2}
\frac{2c\kappa^2(\mu\phi_T-e^{-\mu\phi_T})}{\mu^2}-4\alpha_2(\phi_T)=\ln(\frac{2\lambda\mu^2e^{-2\mu\phi_T}}{cq_0^2\kappa^2})~,
\ee
which may not have an analytical solution.


\subsubsection{Small-field concave inflation}
\begin{figure}[htbp]
\centering
\includegraphics[width=1.70in,height=1.10in]{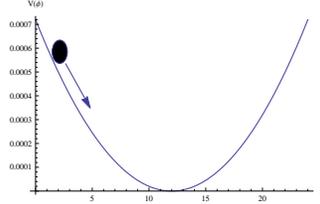}\\
\caption{The plot of $V_3(\phi)=\frac{1}{2}m^2(\phi-\phi_0)^2$.}
\label{v3plot}
\end{figure}

In this case, we have $V'<0$ and $V''>0$ for the potential $V(\phi)$. An example of this potential is:
\be\label{v3}
V_3(\phi)=\frac{1}{2}m^2(\phi-\phi_0)^2~.
\ee
This case corresponds to a shift of the potential used in 3.1.1. Because of this shift, to realize inflation, $\phi$ must begin near zero and end with a large value, generating small-field inflation. The shape of the potential is illustrated in Fig. \ref{v3plot}.

From (\ref{v3}), we have
\be\label{Repsilon3}
\epsilon_{V3}=\frac{2}{\kappa^2(\phi-\phi_0)^2}~,~~~{\cal R}_3=\frac{c\Omega_\phi-1}{c^2\kappa^2\Omega_\phi(\phi-\phi_0)^2}~.
\ee
Because the shift in the potential is such that $\phi$ starts approximately zero, the value of $\phi_0$ is assumed to be roughly the initial value of $\phi$ in the model of (\ref{v1}). Therefore, the efolding number $\alpha$ and the velocity of $\phi$ as functions of $\phi$ are:
\bea\label{alpha3}
\alpha_3(\phi)&=&
\begin{cases}
-\frac{\kappa^2}{4}(\phi-\phi_i)(\phi+\phi_i-2\phi_0)~&(\phi\leq\phi_T)\\
\alpha_3(\phi_T)-\frac{c\kappa^2}{4}(\phi-\phi_T)(\phi+\phi_T-2\phi_0)~&(\phi\geq\phi_T)
\end{cases}~,\\
\label{dotphi3}
\dot\phi(\phi)&=&\sqrt{\frac{2}{3}}\frac{m}{\kappa}~~(\phi\leq\phi_T)~,~\sqrt{\frac{2}{3\Omega_\phi}}\frac{m}{c\kappa}~~(\phi\geq\phi_T)~,
\eea
and $\phi_T$ can be solved as:
\be\label{phiT3}
\phi_{T3}=\phi_0-\sqrt{\phi_0^2+\frac{2\phi_0\phi_i-\phi_i^2}{c-1}-\frac{1}{(c-1)\kappa^2}\ln(\frac{m^2}{cq_0^2\kappa^2})}~.
\ee



\subsubsection{Small-field convex inflation}
\begin{figure}
\centering
\includegraphics[width=1.70in,height=1.10in]{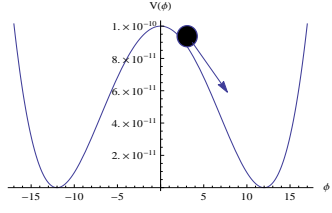}\\
\caption{The plot of $V_4(\phi)=V_0\left(1-\frac{\phi^2}{\mu^2}\right)^2$.}
\label{v4plot}
\end{figure}

In the last case, we have $V'<0$ and $V''<0$ for the potential $V(\phi)$. There are many famous examples of this case; for example, the potential can be  of the form:
\be\label{v4}
V_4(\phi)=V_0\left(1-\frac{\phi^2}{\mu^2}\right)^2~,
\ee
which can be applied to symmetry-breaking inflation models such as the Higgs inflation model \cite{Bezrukov:2007ep}. The shape of the potential is shown in Fig. \ref{v4plot}.

From (\ref{v4}), we have
\be\label{Repsilon4}
\epsilon_{V4}=\frac{8\phi^2}{\kappa^2(\mu^2-\phi^2)^2}~,~~~{\cal R}_4=\frac{4(c\Omega_\phi-1)\phi^2}{c^2\kappa^2\Omega_\phi(\mu^2-\phi^2)^2}~,
\ee
and the efolding number $\alpha$ and the velocity of $\phi$ as functions of $\phi$ are:
\bea\label{alpha4}
\alpha_4(\phi)&=&
\begin{cases}
\frac{\kappa^2}{8}[\phi_i^2-\phi^2+2\mu\ln(\frac{\phi}{\phi_i})]~&(\phi\leq\phi_T)\\
\alpha_4(\phi_T)+\frac{c\kappa^2}{8}[\phi_T^2-\phi^2+2\mu\ln(\frac{\phi}{\phi_T})]~&(\phi\geq\phi_T)
\end{cases}~,\\
\label{dotphi4}
\dot\phi(\phi)&=&\frac{4}{\mu^2\kappa}\sqrt{\frac{V_0}{3}}\phi~~(\phi\leq\phi_T)~,~\frac{4}{c\mu^2\kappa}\sqrt{\frac{V_0}{3\Omega_\phi}}\phi~~(\phi\geq\phi_T)~,
\eea
where $\phi_T$ is the solution of the equation:
\be\label{phiT4}
c\kappa^2(\mu^2\ln\phi_T-\phi_T^2/2)-4\alpha_4(\phi_T)=\ln(\frac{8V_0\phi_T^2}{cq_0^2\kappa^2\mu^4})~,
\ee
which may not have an analytical solution.



\subsection{Numerical calculations}
\begin{figure}
\centering
\includegraphics[width=1.70in,height=1.30in]{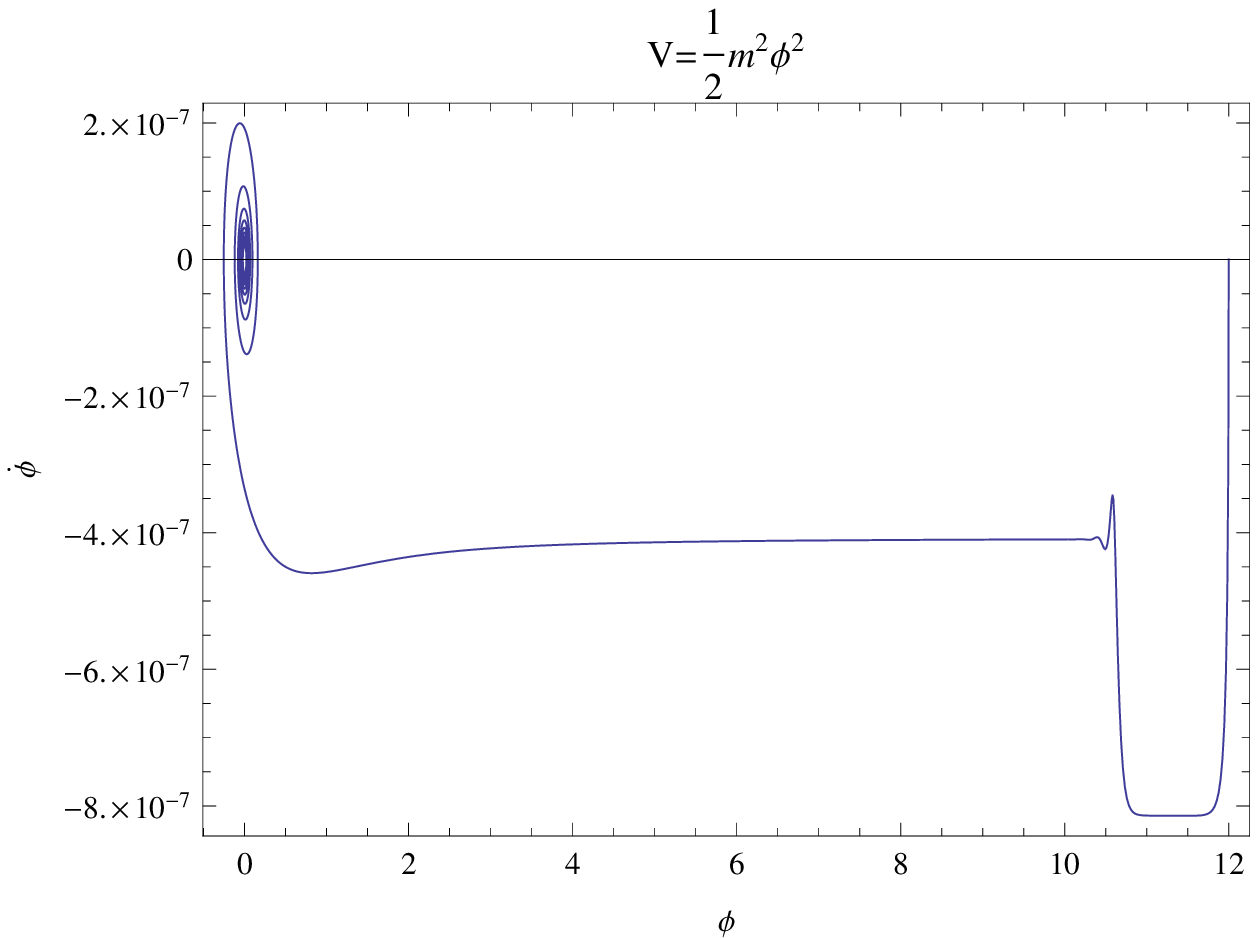}
\includegraphics[width=1.70in,height=1.30in]{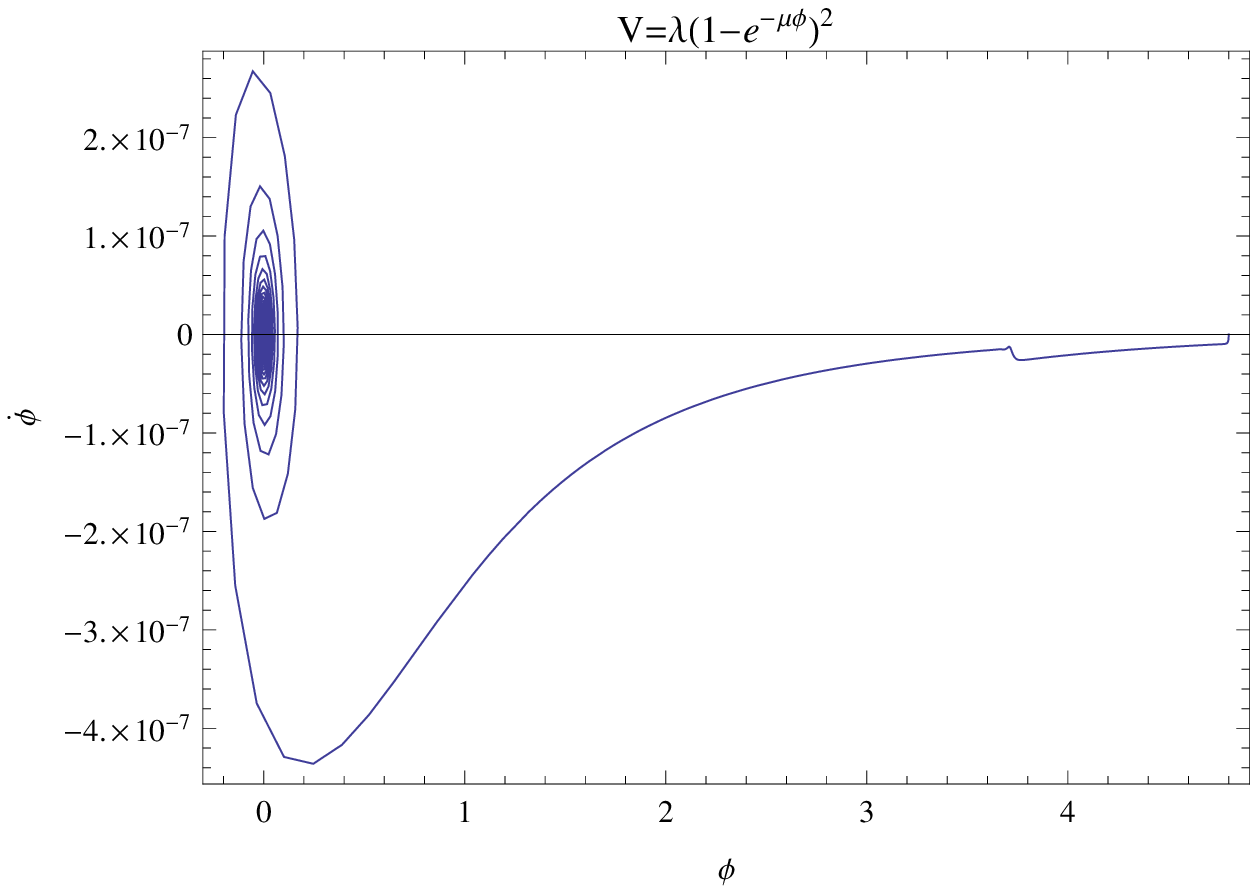}\\
\includegraphics[width=1.70in,height=1.30in]{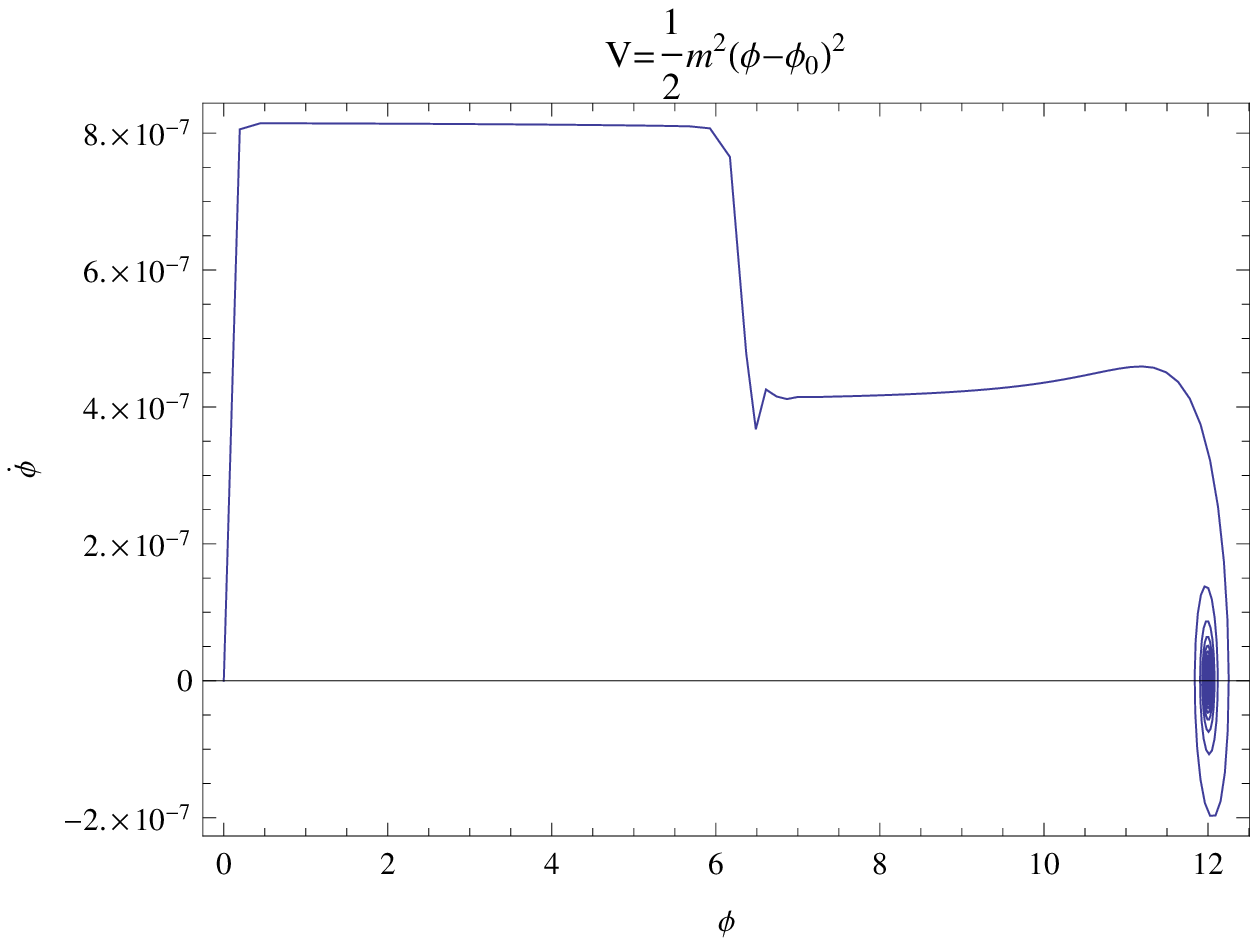}
\includegraphics[width=1.70in,height=1.30in]{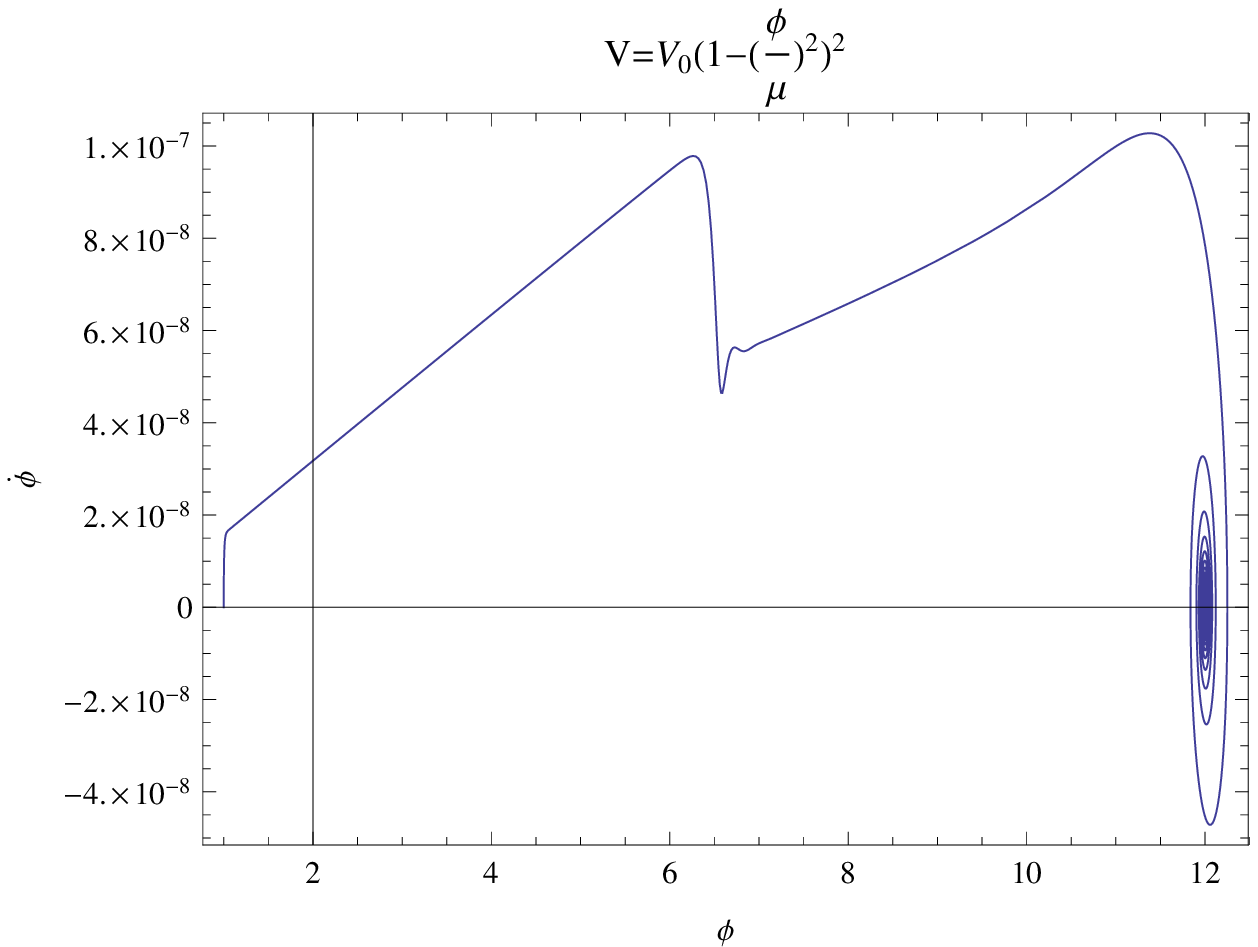}\\
\caption{The plots of $\phi-\dot\phi$ during inflation in the four cases. Here the parameters were selected as: $m=10^{-6}M_p$, $\phi_i=12M_p$ for $V_1$; $\lambda=10^{-12}M_p^4$, $\mu=1$, $\phi_i=4.8M_p$ for $V_2$; $m=10^{-6}M_p$, $\phi_0=12M_p$, $\phi_i=0$ for $V_3$; $V_0=10^{-12}M_p^4$, $\mu=12M_p$, $\phi_i=1M_p$ for $V_4$. For all cases we choose $c=2$ for the vector field. }
\label{pp}
\end{figure}

In the following, we perform a numerical calculation for each case to study the behavior of our model. First, we calculate the parameter space of $(\phi,\dot\phi)$. By using basic equations (\ref{phieom}), (\ref{Fr1}), (\ref{Fr2}) and (\ref{ani}), we obtain four plots of $\phi-\dot\phi$ for each case in Fig. \ref{pp}. The plots demonstrate the phase transition occurs in all cases, after which the vector field plays a certain role. In the upper plots (large-field cases), we have $\dot\phi<0$ during inflation, while in the lower plots (small-field cases), $\dot\phi>0$. Moreover, while in the left plots (concave model), $\dot\phi(\phi)$ remains mostly constant, in the right plots (convex model), $\dot\phi$ behaves as a monotonic function of $\phi$, which is consistent with our analytical solutions (\ref{dotphi1}), (\ref{dotphi2}), (\ref{dotphi3}) and (\ref{dotphi4}).

\begin{figure}[htbp]
\centering
\includegraphics[width=2.00in,height=1.50in]{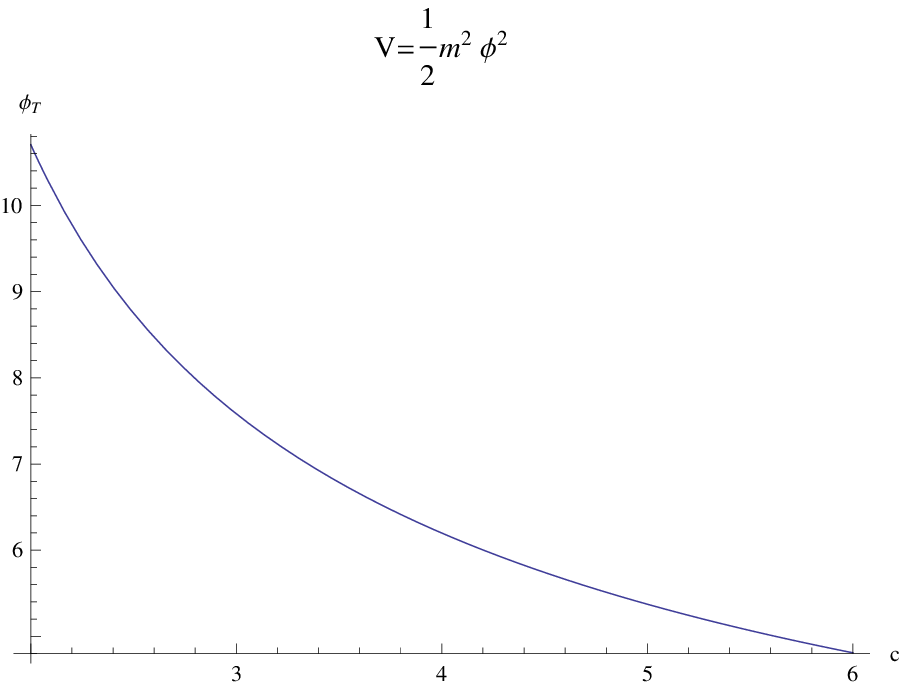}
\includegraphics[width=2.00in,height=1.50in]{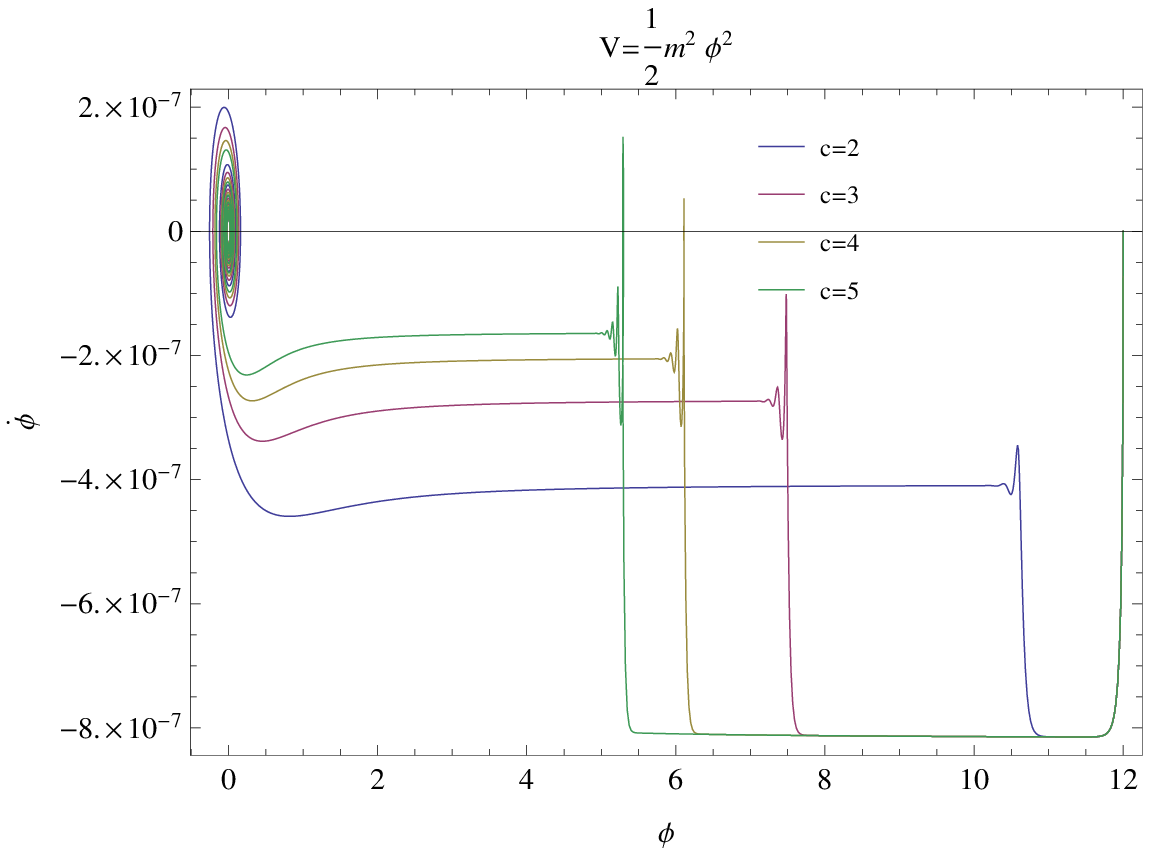}\\
\includegraphics[width=2.00in,height=1.50in]{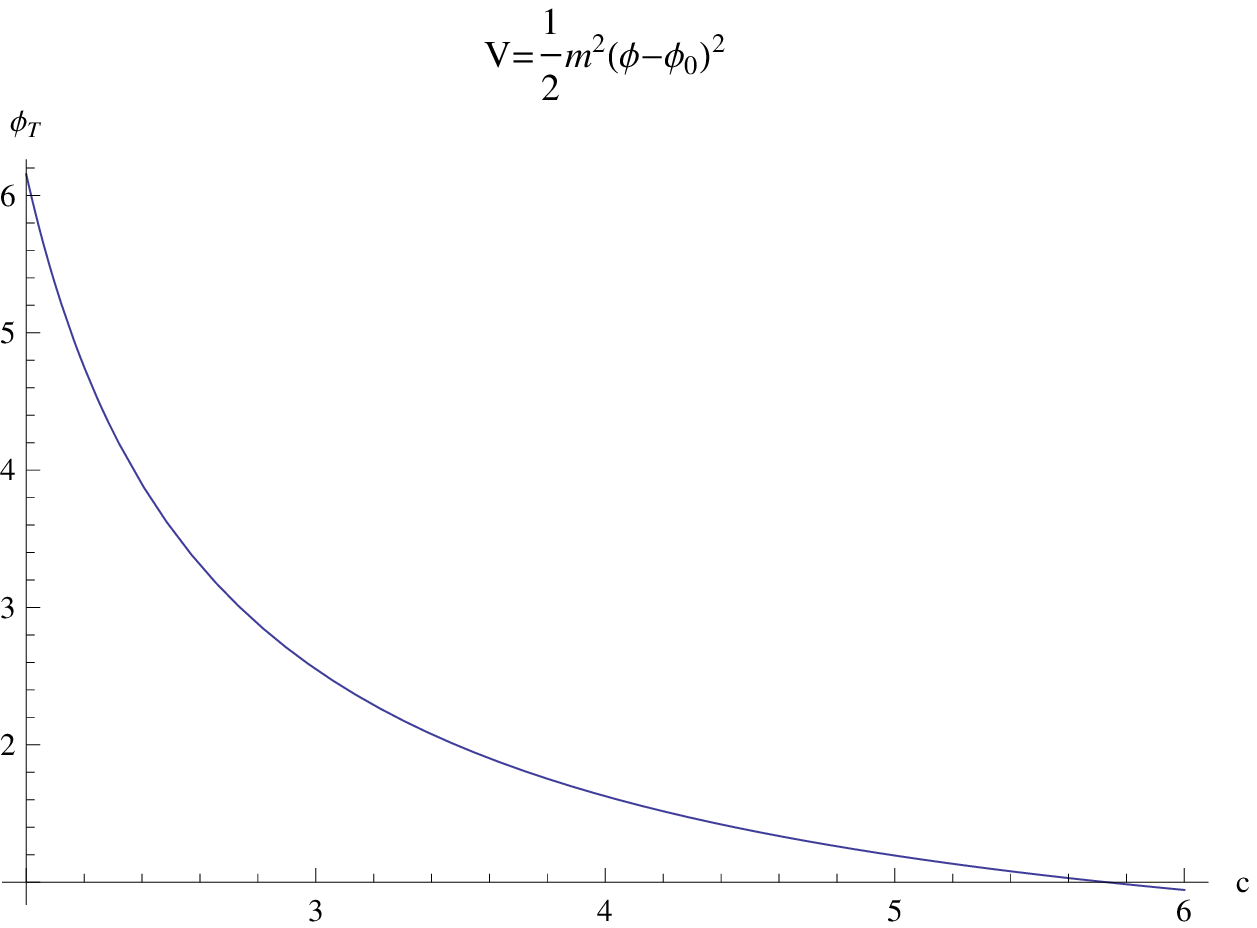}
\includegraphics[width=2.00in,height=1.50in]{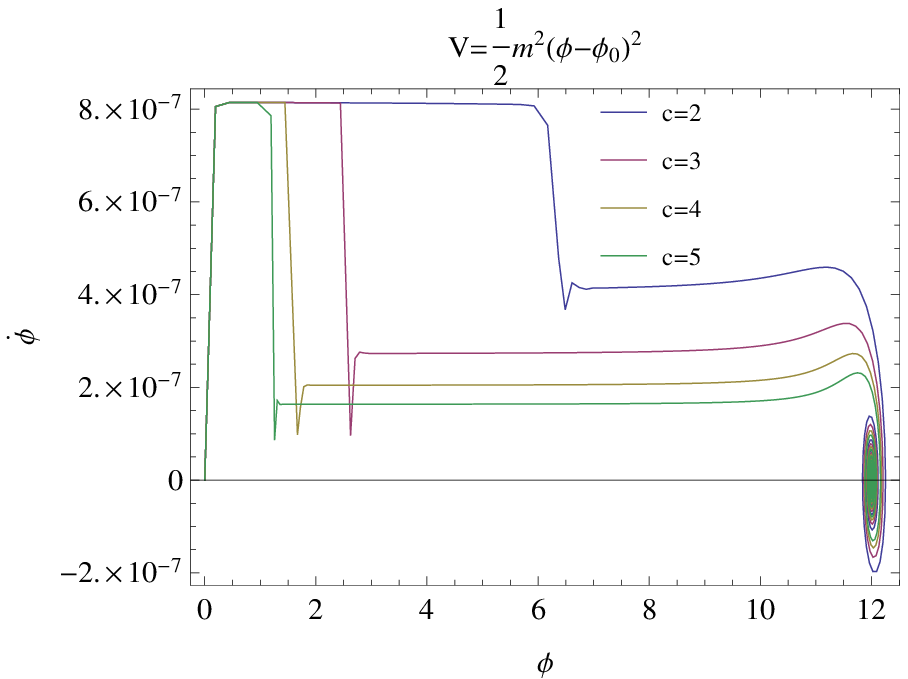}\\
\caption{{\it Left:} The relationship between $\phi_T$ and $c$ for the large- and small-field cases. {\it Right:} The plots of $\phi-\dot\phi$ during inflation for different values of $c$ in the large- and small-field cases.}
\label{phiTcplot}
\end{figure}

Subsequently, we consider how the phase transition depends on the parameter $c$. Because there are no analytical solutions for $\phi_T$ in the second and fourth cases, we have numerically plotted (in Fig. \ref{phiTcplot}) $\phi_T$ as a function of $c$ only for the first and third cases, which represent large-field models and small-field models, respectively. In the plots, we can see that in both cases, $\phi_T$ decreases with respect to $c$; this, however, leads to quite different consequences for the different models. For large-field models, where $\phi$ evolves from large value to small value, it means that if we increase $c$ and thus suppress the value of $\phi_T$, the phase transition will be postponed to a later time. In contrast, for small-field models, where $\phi$ increases from a small to a large value, if we increase $c$ and thus suppress the value of $\phi_T$, the phase transition will be promoted. The difference could generally be explained as follows: from Eq. (\ref{Fr1}), the strength of the coupling between the vector and scalar fields can be considered as proportional to $f^{-2}(\phi)$. For large-field models, where $V'>0$, a larger $c$ results in a larger $f(\phi)$, which suggests weaker coupling, and the vector field obtains sufficient energy from the scalar to play an important role at a later time. However, for small-field models, where $V'<0$, the opposite occurs.

\begin{figure}[htbp]
\centering
\includegraphics[width=2.0in,height=1.50in]{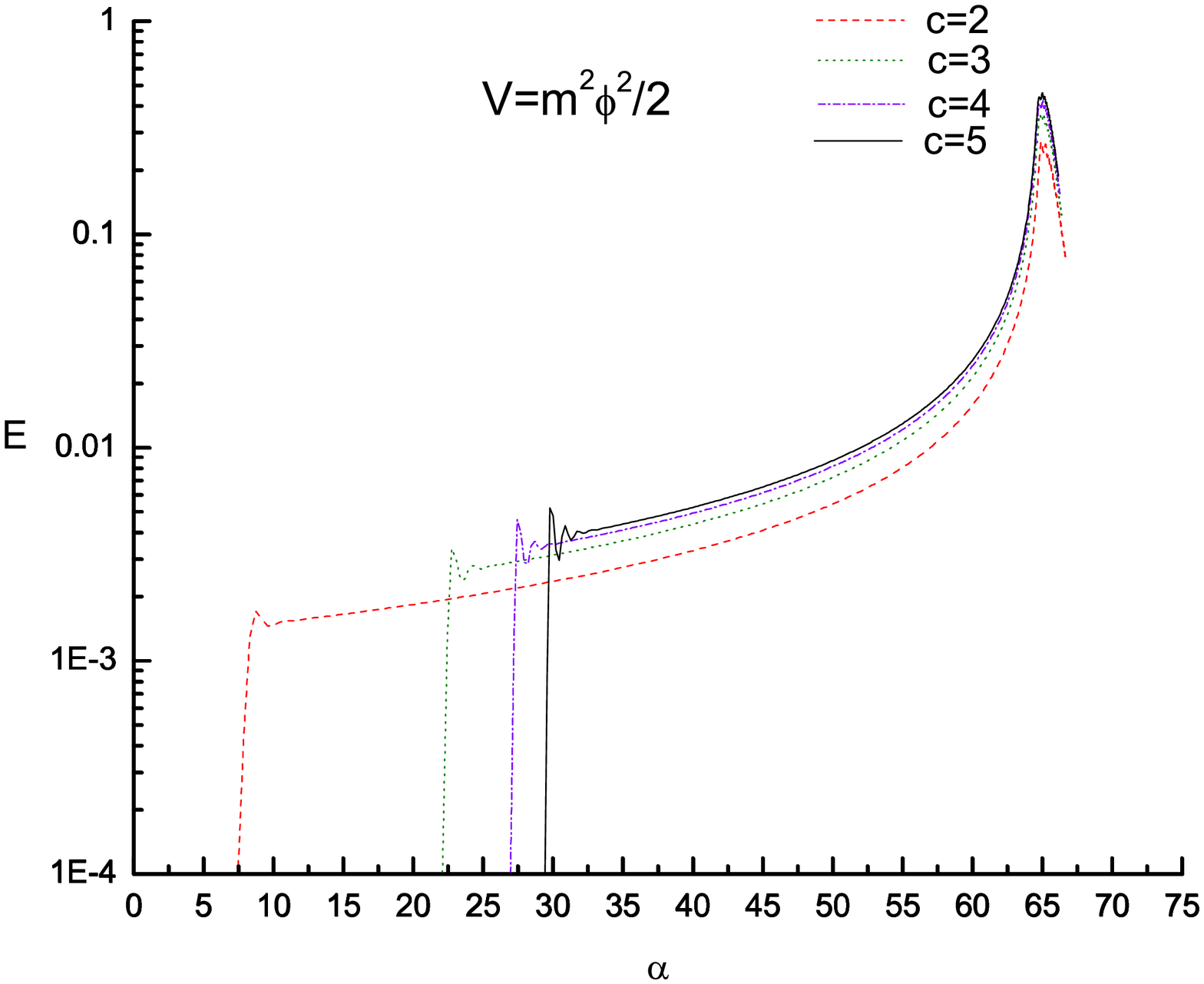}
\includegraphics[width=2.0in,height=1.50in]{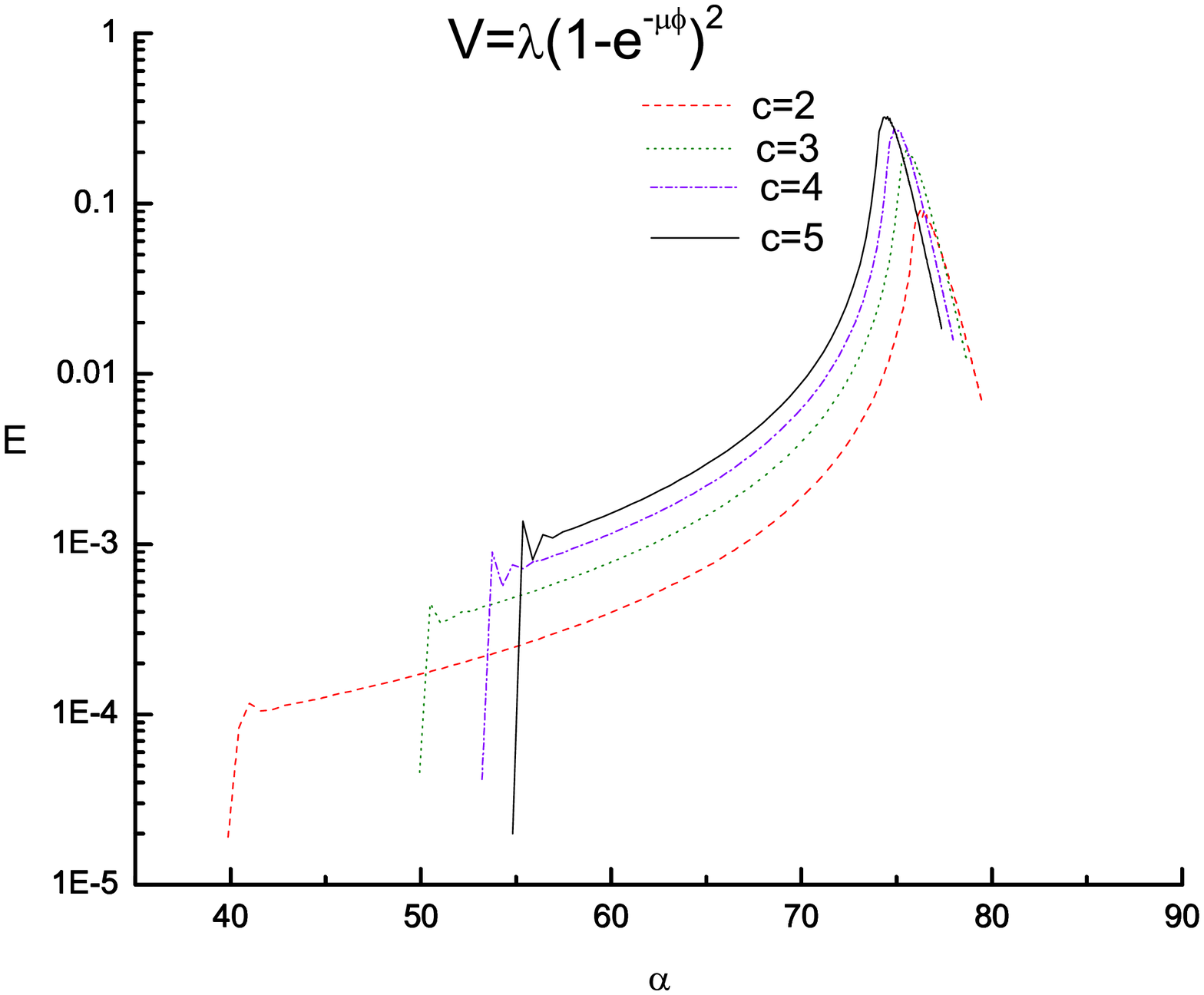}\\
\includegraphics[width=2.0in,height=1.50in]{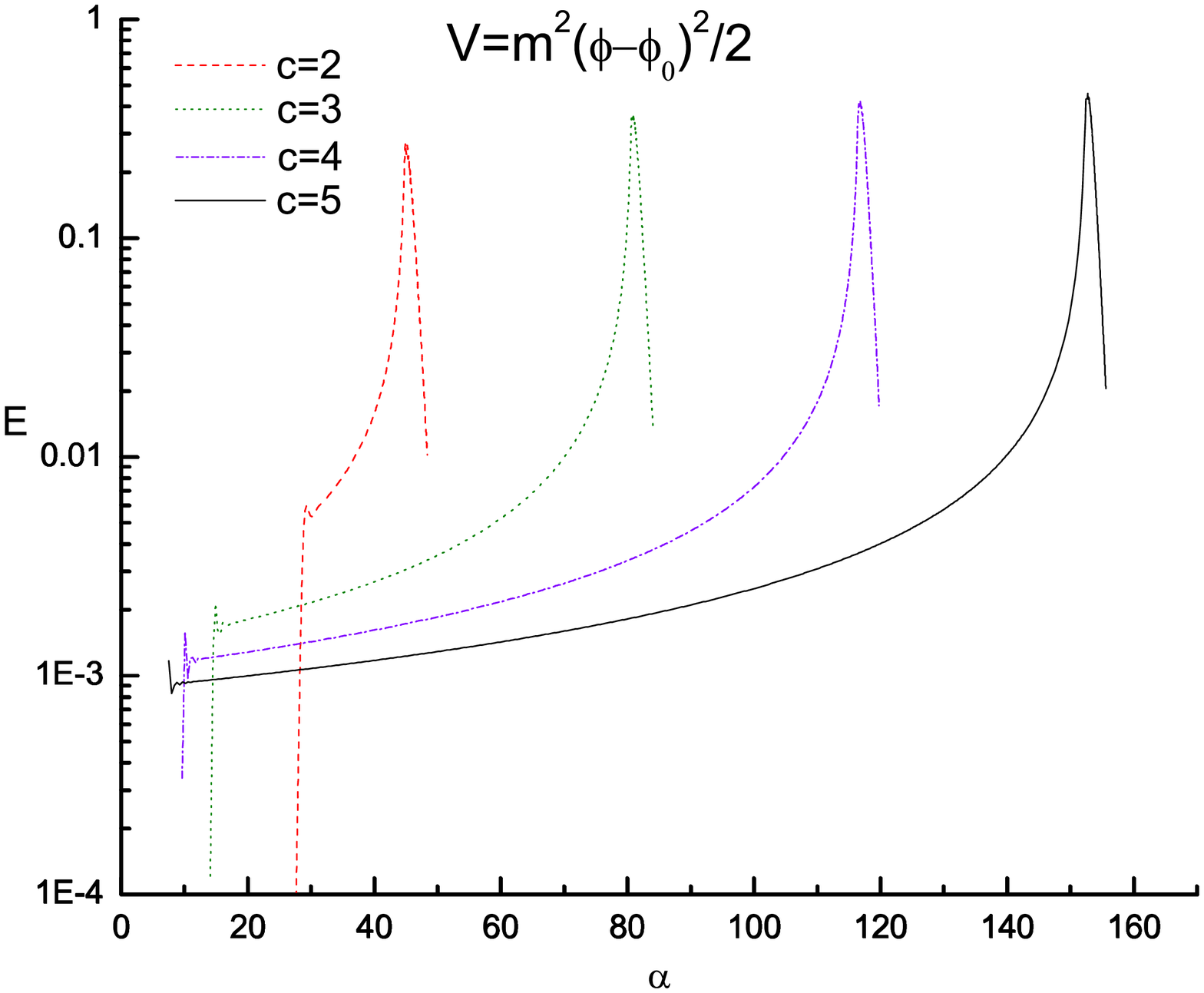}
\includegraphics[width=2.0in,height=1.50in]{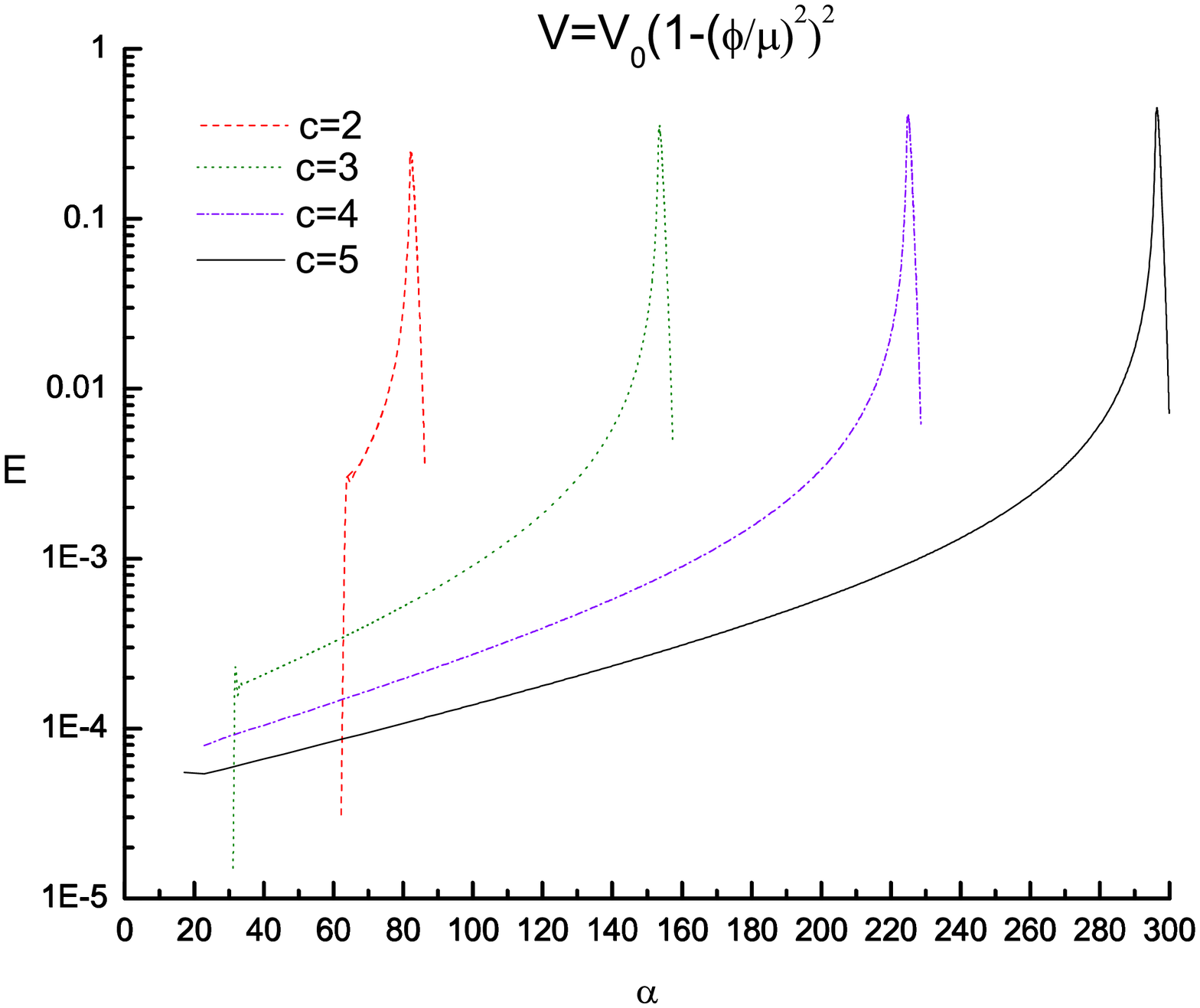}\\
\caption{The evolution of ${\cal E}$ with respect to the efolding number $\alpha$ in the four examined cases.}
\label{Ealphaplot}
\end{figure}

\begin{figure}[htbp]
\centering
\includegraphics[width=2.0in,height=1.50in]{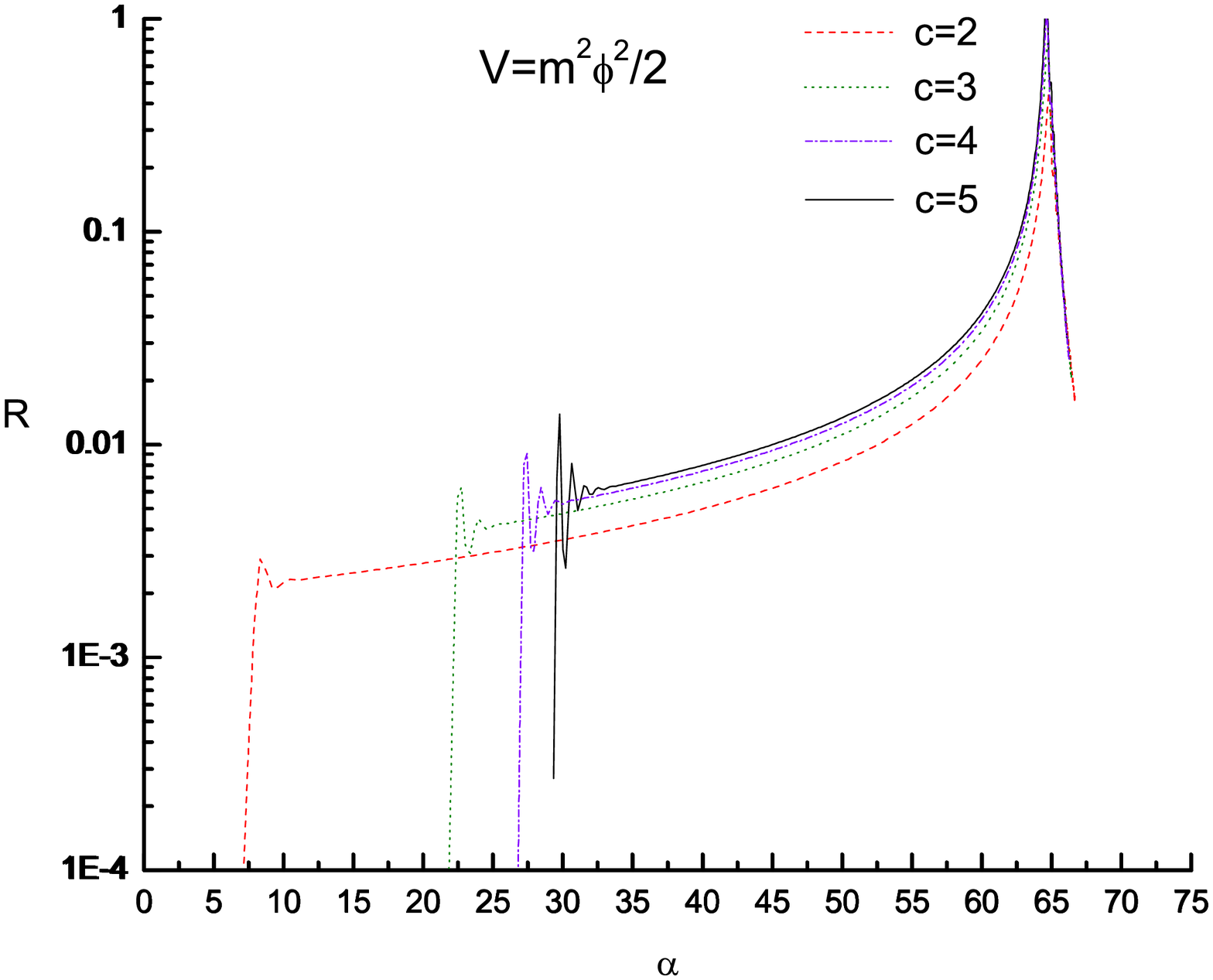}
\includegraphics[width=2.0in,height=1.50in]{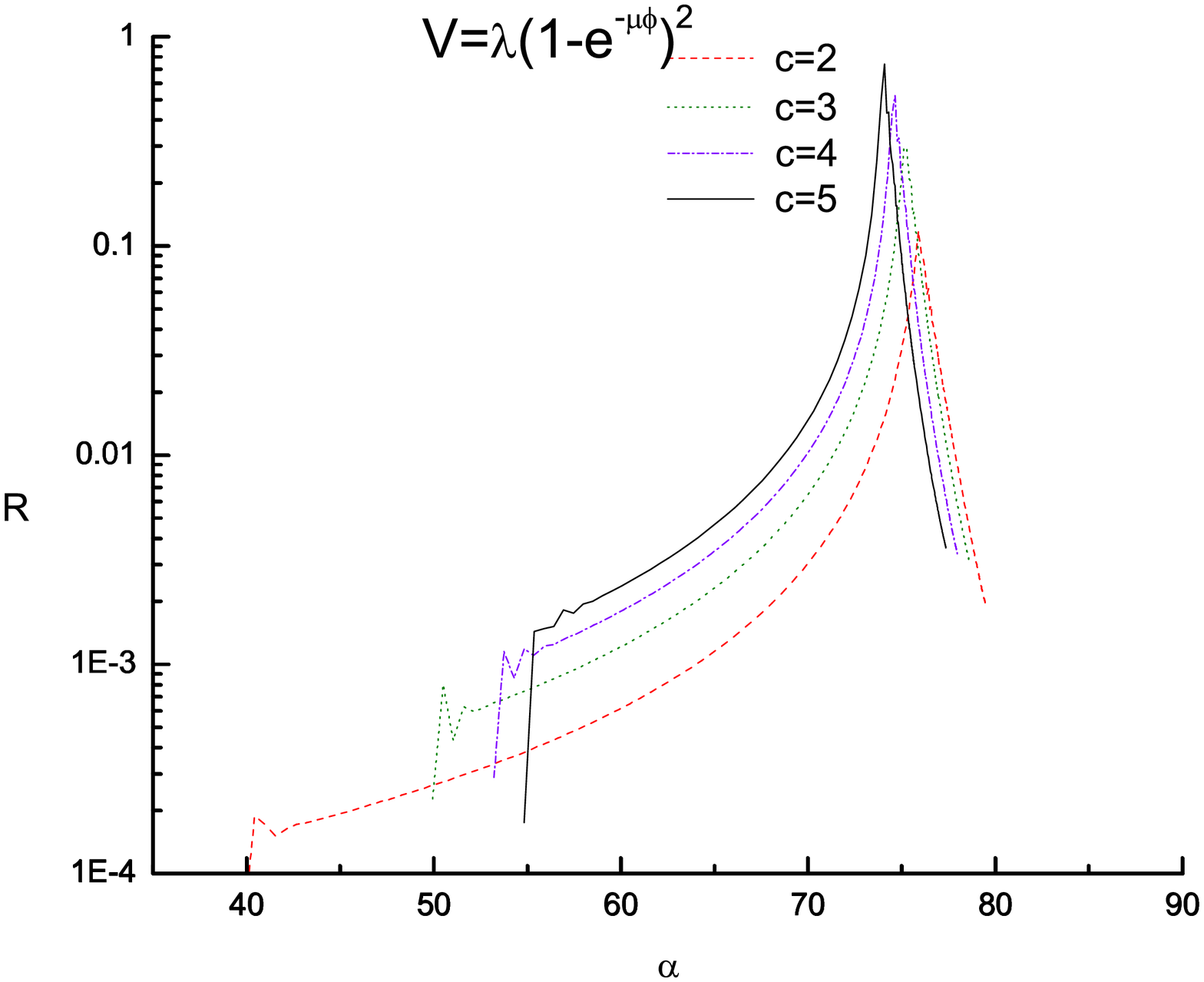}\\
\includegraphics[width=2.0in,height=1.50in]{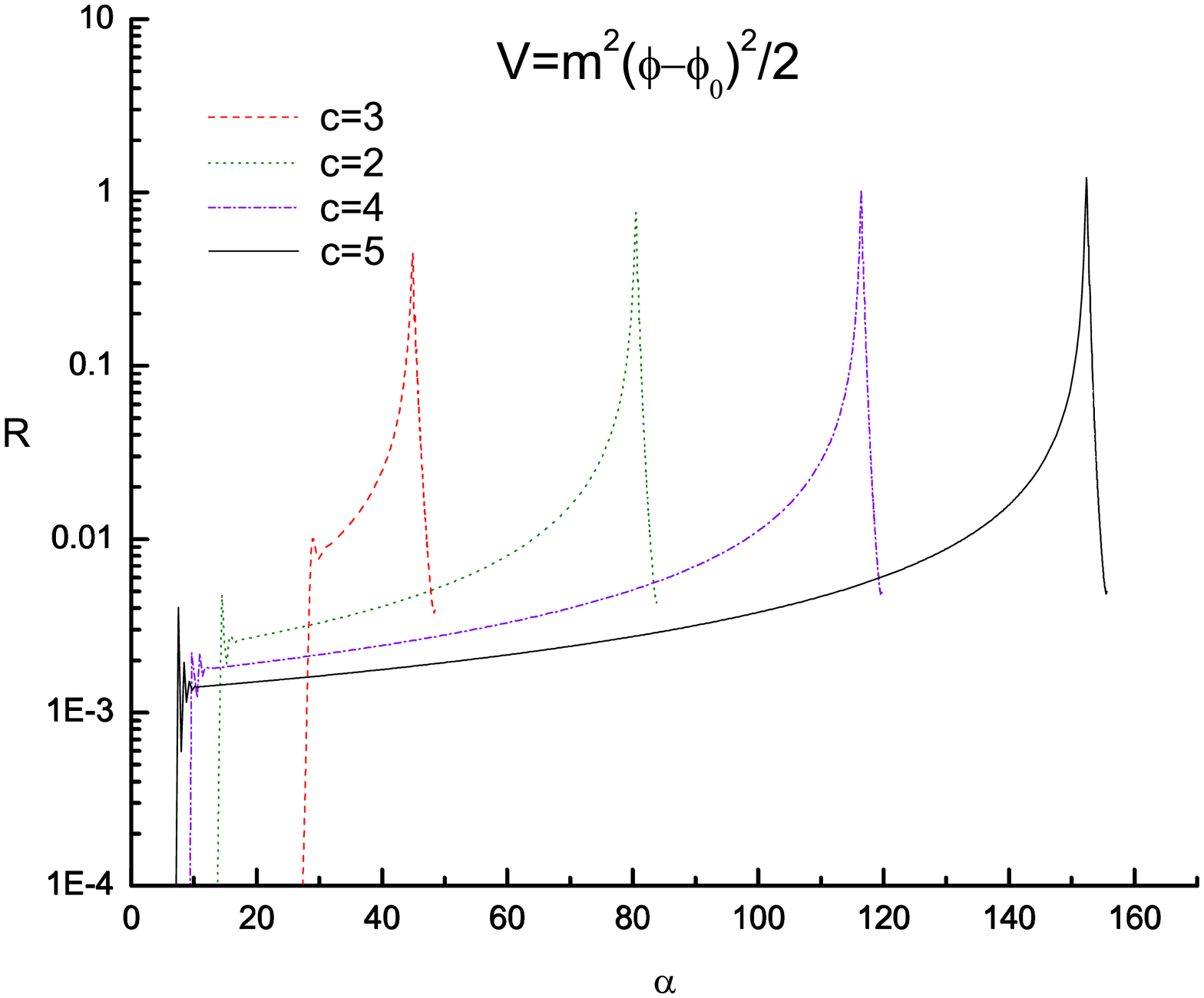}
\includegraphics[width=2.0in,height=1.50in]{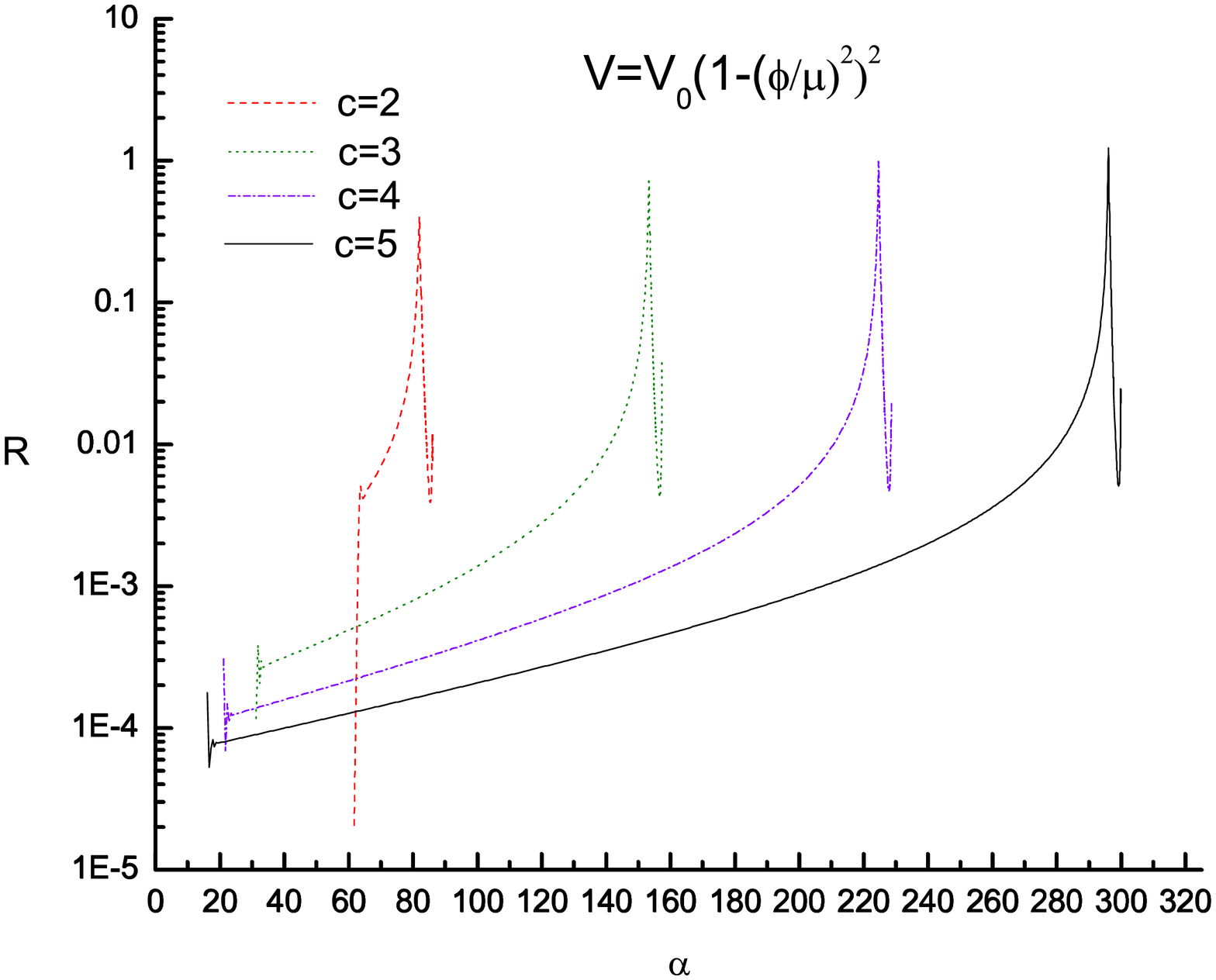}\\
\caption{The evolution of ${\cal R}$ with respect to the efolding number $\alpha$ in the four examined cases.}
\label{Ralphaplot}
\end{figure}

Finally, we consider the evolutions of ${\cal R}$ and ${\cal E}$ in the four cases. In Fig. \ref{Ealphaplot}, we have plotted the evolution of ${\cal E}$ with respect to the efolding number $\alpha$. It is evident that all four cases exhibit similar behavior. This indicates that the evolution of anisotropic inflation has an attractor solution and is independent of the form of the potentials. Moreover, we can see that in all four cases, ${\cal E}$ varies between ${\cal O}(10^{-5})$ and ${\cal O}(1)$ after the vector field has been considered; this suggests that the anisotropy plays a certain role but does not break the global isotropy of the background. We also observe that in the large-field case, a larger $c$ results in a later the phase transition, while the opposite occurs in the small-field case, which is consistent with the result shown in Fig. \ref{phiTcplot}. In Fig. \ref{Ralphaplot}, which displays the evolution of ${\cal R}$ with respect to the efolding number $\alpha$, we can see that $R$ also varies in the same range as ${\cal E}$.

\section{Extension: Multi-Directional Anisotropies}
In the previous sections, we mainly focused on the hypothesis that the anisotropies are normalized in one direction. It is therefore reasonable to extend the analysis to multi-directed anisotropies, in which the cosmic expansions are different in all the three spatial directions. To describe this case, the metric (\ref{metric}) can be extended to the form:
\be\label{metric2}
ds^2=-dt^2+e^{2\alpha}\left[e^{2\sigma_1}dx^2+dy^2+e^{-2\sigma_2}dz^2\right]~,
\ee
where the anisotropies have two degrees of freedom, $\sigma_1$ and $\sigma_2$. For the vector field, again we choose $A_\mu=(0,A_x(t),0,0)$ so that the anisotropy of the $x$-direction is sourced by $A_x$ while that of the $z$-direction is not sourced by anything. From the action (\ref{action}), we determine a new equation of motion for $A_\mu$:
\be\label{Aeom2}
\dot{A_x}=q_0f^{-2}(\phi)e^{-\alpha+\sigma_1+\sigma_2}~.
\ee
The Friedmann equation and the equation of motion for $\sigma_1$ and $\sigma_2$ are:
\bea
\dot\alpha^2&=&\frac{2}{3}\dot{\alpha}(\dot{\sigma_2}-\dot{\sigma_1})+\frac{1}{3}\dot{\sigma_1}\dot{\sigma_2}
+\frac{\kappa^2}{3}\left[\frac{1}{2}\dot{\phi}^2+V(\phi)+\frac{1}{2}q_0^2f^{-2}(\phi)e^{2\sigma_2-4\alpha}\right]~,\\
\ddot\sigma_1&=&-3\dot{\alpha}\dot{\sigma_1}-\dot{\sigma_1}^2+\dot{\sigma_1}\dot{\sigma_2}-\kappa^2q_0^2f^{-2}(\phi)e^{2\sigma_2-4\alpha}~,\\
\ddot\sigma_2&=&-3\dot{\alpha}\dot{\sigma_2}+\dot{\sigma_2}^2-\dot{\sigma_1}\dot{\sigma_2}~.
\eea
We can also define the degree of anisotropy for both $x$ and $z$ directions, which is:
\be
{\cal E}_x\equiv\frac{\dot\sigma_1}{\dot\alpha}~,~~~{\cal E}_z\equiv\frac{\dot\sigma_2}{\dot\alpha}~.
\ee

\begin{figure}
\centering
\includegraphics[width=1.7in,height=1.30in]{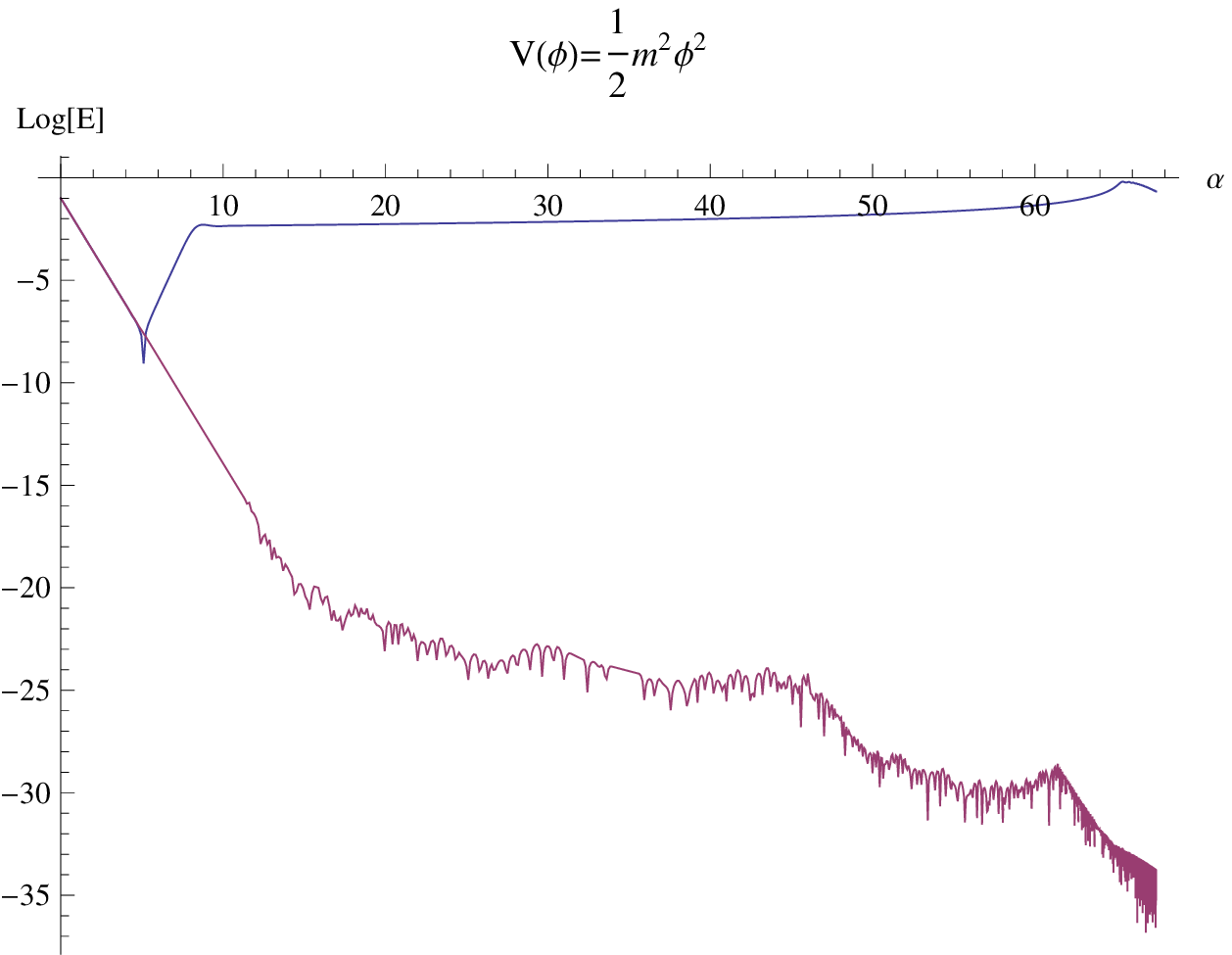}
\includegraphics[width=1.7in,height=1.30in]{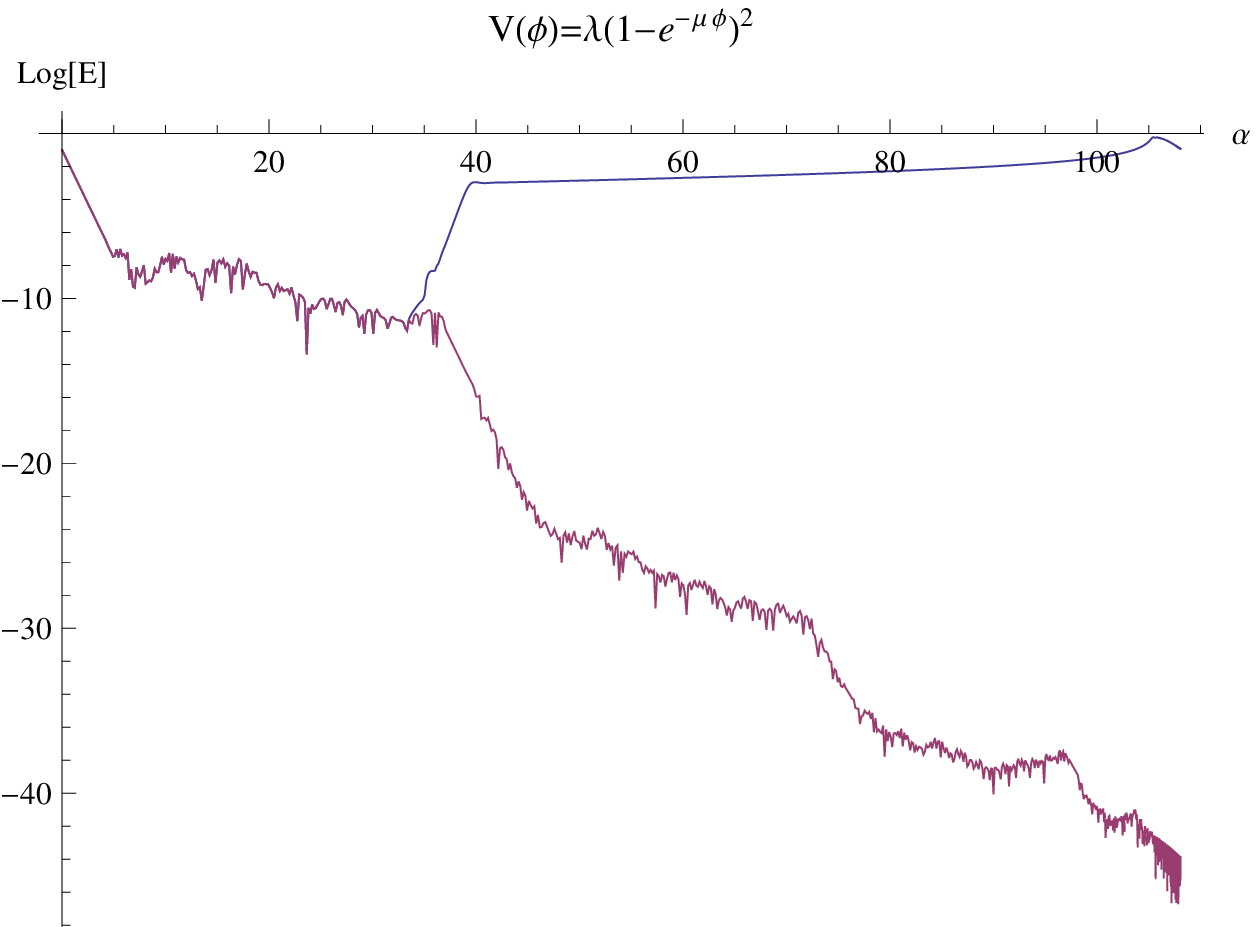}\\
\includegraphics[width=1.7in,height=1.30in]{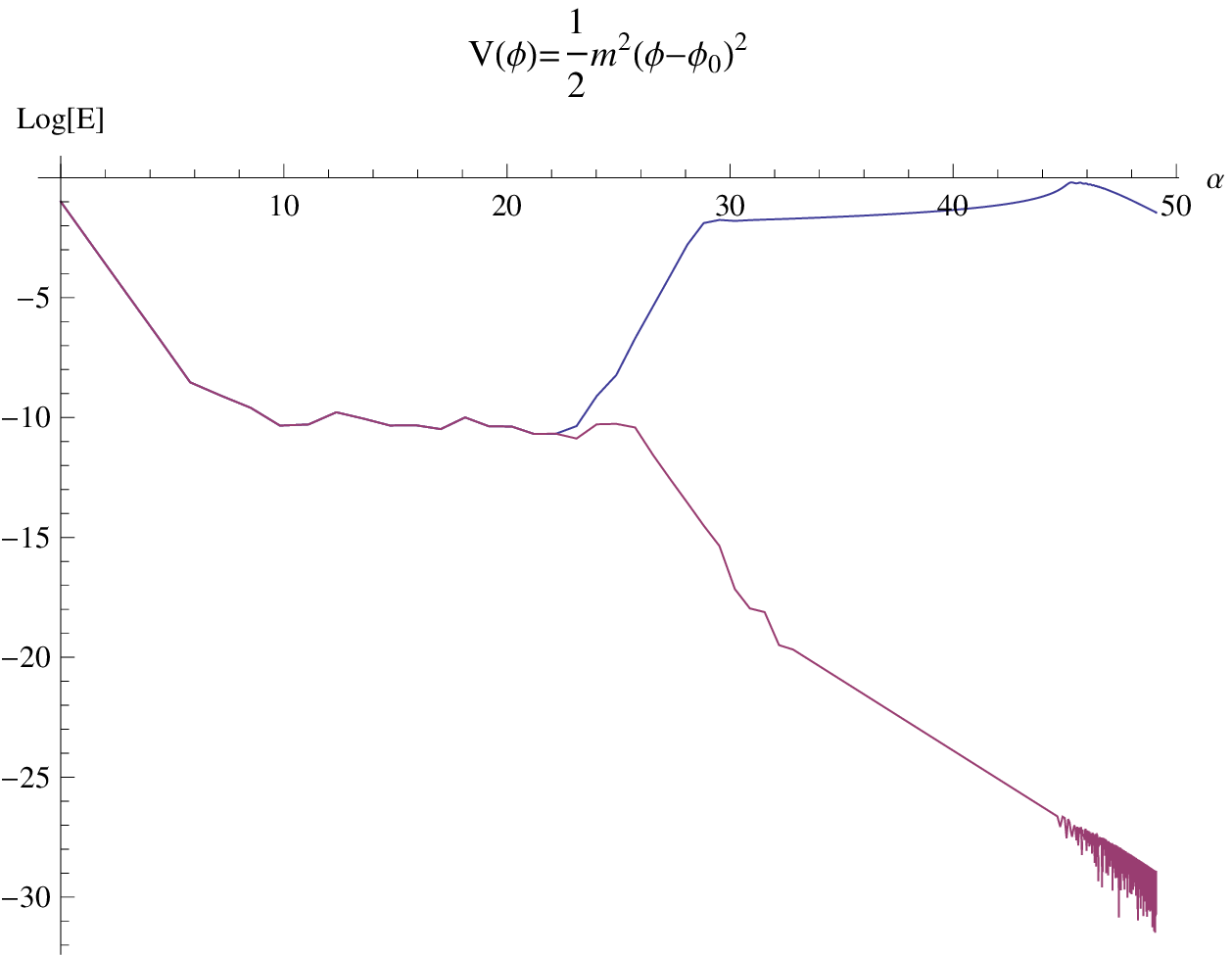}
\includegraphics[width=1.7in,height=1.30in]{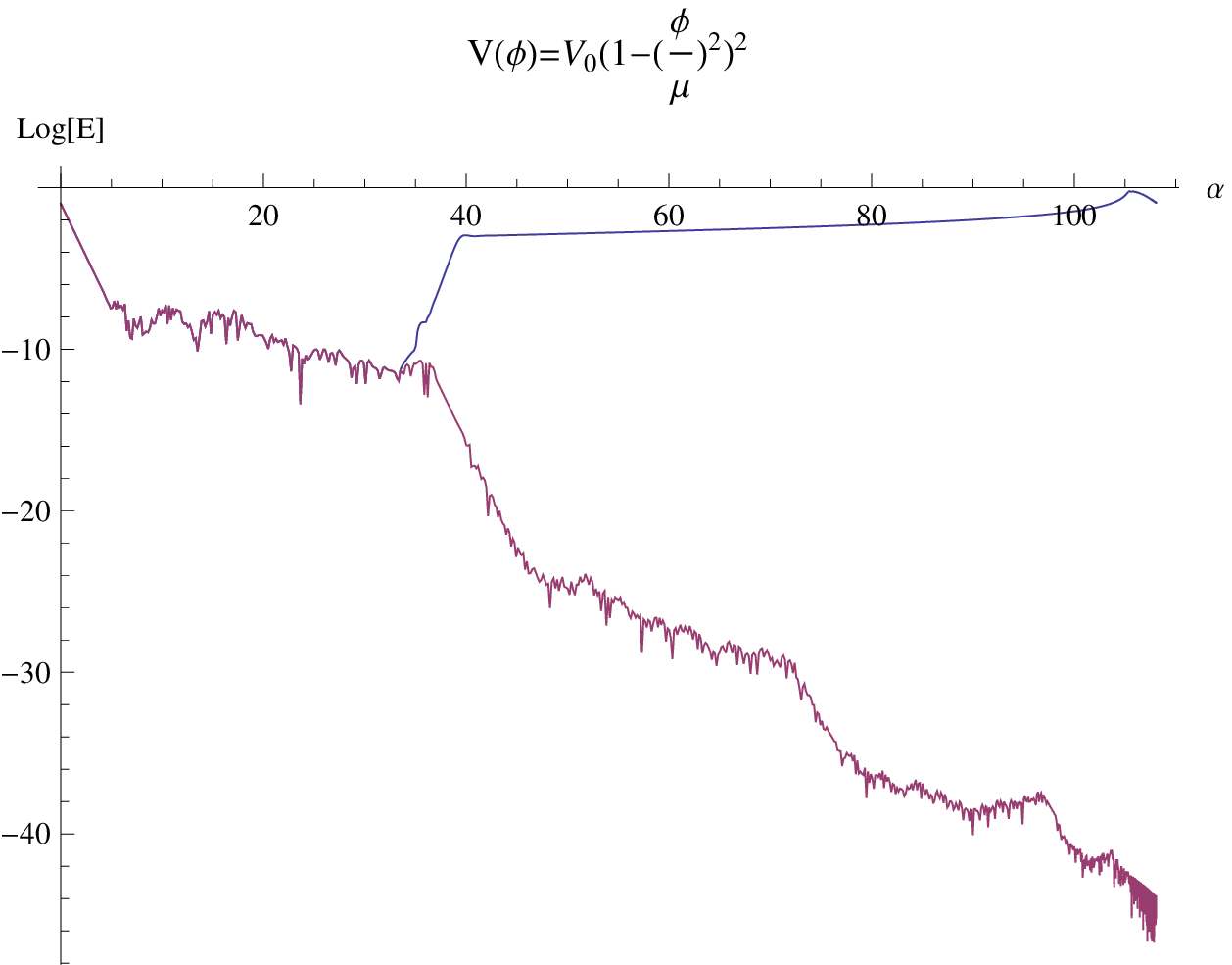}\\
\caption{Plots of $\log{\cal E}_x$ (blue) and $\log{\cal E}_z$ (red) with respect to the e-folding number $\alpha$. In all four cases, vector inflation enhances ${\cal E}_x$ and weakens ${\cal E}_z$.}
\label{ani2}
\end{figure}

In Fig. \ref{ani2}, we  have numerically plotted $\log{\cal E}_x$ and $\log{\cal E}_z$ for the four cases discussed in Sec. 3. for initial conditions  $\dot{\sigma_1}/\dot{\alpha}|_{t=0}=\dot{\sigma_2}/\dot{\alpha}|_{t=0}=0.1$. In the plot, we observe that the anisotropies in the two directions evolve in significantly different ways. The sourced anisotropy ${\cal E}_x$ exhibits an increasing behavior whereas the unsourced anisotropy ${\cal E}_z$ continuously decays.

We can also extend the analysis to the case in which the anisotropies are sourced in both directions, which requires the vector field to be of the form $A_\mu=(0,A_x(t),0,A_z(t))$. In this case, the same numerical study determines that both anisotropies are enhanced. This result means that the vector source only enhances the anisotropies in the same direction but not those in other directions.

\section{Conclusion}
The anomalies on the CMB map described by the PLANCK data indicate that anisotropic hair may have been generated in the early universe. This anisotropic hair can be naturally modeled by a vector field coupled to the inflation field. As observations become increasingly accurate, the precise study of the properties of such "anisotropic inflation" becomes more interesting.

As an extension of Ref. \cite{Watanabe:2009ct} and many other works, we revisited the anisotropic inflation models, and obtained some interesting findings. Firstly, in the analytical calculations of most previous works, the vector field and anisotropy term $\dot\sigma^2$ were usually neglected in the calculation of the total energy density. Although these terms are subdominant, by keeping them in the equations we determined that the linear relationship between the degree of anisotropy ${\cal E}$ and the energy density fraction ${\cal R}$ no longer holds, as in previous works, and a nonlinear correction appears. As demonstrated in this paper, this imposes an upper limit on the degree of anisotropy, namely ${\cal E}_{max}\simeq 0.5$. This means that the inflation is not ruined even in the case of vector field domination. However, because ${\cal E}$ exceeds $o(0.1)$ in this case, care must be taken regarding its effects on the total energy density. The same analysis was performed for ${\cal R}$ and the slow-roll parameters $\epsilon_V$ and $\epsilon_\alpha$. In the regions where they become large (near the end of inflation), the linear relationship between ${\cal R}$ and the slow-roll parameter is altered.

Secondly, we investigated various forms of inflaton potentials, including large/small-field models and concave/convex-field models. We observed that qualitatively, the effects of the vector field are almost the same for all four cases (large-concave, large-convex, small-concave, and small-convex). The vector becomes important for the later period of inflation but does not ruin the inflation. However, there are slight quantitative differences. For large-field models, a larger parameter $c$ results in a later phase transition (after which the vector begins to play an important role) whereas for small-field models, the opposite occurs. This might be owing to the fact that they have opposite signs of $V^\prime$ (which appears in the exponential index of the coupling function $f(\phi)$ between scalar and vector), which will have the opposite effects on the energy transfer between them to make the vector important. For concave and convex types, however, there are no explicit differences.

Finally, we investigated the case of multi-directional anisotropies, i.e., the expansion rates are different in the three spatial dimensions. We determined that the anisotropy increases in the dimensions that are sourced by the vector field whereas it decays in those that are not sourced. This means that the vector field does not source the anisotropies in a different dimension, although it appears to have couplings between different dimensions.

\section*{Acknowledgements}
This work is supported by the Project of Undergraduates' Training Program for Innovation/Venturing, Central China Normal University, under the No. B2014179.


\end{document}